\documentclass[aps,pra,floatfix,longbibliography, superscriptaddress,twocolumn]{revtex4-2}
\usepackage{xcolor}
\usepackage{appendix}
\usepackage{blindtext}
\newcommand*{\colorboxed}{}
\def\colorboxed#1#{%
	\colorboxedAux{#1}%
}
\newcommand*{\colorboxedAux}[3]{%
	\begingroup
	\colorlet{cb@saved}{.}%
	\color#1{#2}%
	\boxed{%
		\color{cb@saved}%
		#3%
	}%
	\endgroup
}

\usepackage{pifont}

\usepackage{float}
\usepackage{epsfig}
\usepackage{bm}
\usepackage[matrix,frame,arrow]{xy}
\usepackage[applemac]{inputenc}
\usepackage[T1]{fontenc}
\usepackage{lmodern}
\usepackage[english]{babel}
\usepackage{amsmath}
\usepackage{ae}
\usepackage{amssymb}
\usepackage{color}
\usepackage{graphicx}
\usepackage{bbm}
\usepackage{url}
\usepackage{nomencl}
\usepackage{subfigure}
\usepackage{slashed}

\usepackage{booktabs}

\newcommand{\brakkets}[3]{\langle #1| #2 | #3\rangle}

\newcommand{\expec}[1]{\left\langle #1 \right\rangle}

\newcommand{\comm}[2]{\left[ #1, #2 \right]}
\newcommand{\lind}[1]{\mathcal{D}\!\left[#1\right]}

\newcommand{\abs}[1]{\left|#1\right|}

\newcommand{\abssq}[1]{\left| #1 \right|^2}

\newcommand{\figref}[1]{\mbox{Fig.~\ref{#1}}}

\newcommand{\be}{\begin{equation}}
\newcommand{\ee}{\end{equation}}
\newcommand{\bea}{\begin{eqnarray}}
\newcommand{\eea}{\end{eqnarray}}

\usepackage{xr}

\usepackage[colorlinks]{hyperref}
\hypersetup{%
	plainpages=true,
	pdfpagemode={UseOutlines},
	pdfstartview={FitH},
	breaklinks=true,
	hypertexnames=false,
	pageanchor=true,
	colorlinks=true,
	linkcolor={blue},
	citecolor={red},
	urlcolor={blue},
	anchorcolor={black}
}

\newcommand{\beq}{\begin{eqnarray}}
\newcommand{\eeq}{\end{eqnarray}}

\makeatletter

\usepackage{etoolbox}

\newcounter{eqncount}
\newcommand{\eqsRef}[1]{%
	\setcounter{eqncount}{0}
	\renewcommand*{\do}[1]{\stepcounter{eqncount}}
	\docsvlist{#1}
	Eq%
	\ifnum\value{eqncount}>1
	s
	\fi
	.\nobreakspace
	\def\eqnrefdelim{\def\eqnrefdelim{), (}}%
	\renewcommand{\do}{\eqnrefdelim\ref}%
	\textup{[(}%
	\docsvlist{#1}%
	\textup{)]}%
}

\usepackage{hyperref}
\hypersetup{%
    plainpages=true,
    breaklinks=true,
    hypertexnames=false,
    pageanchor=true,
    colorlinks=true,
    linkcolor={blue},
    citecolor={red},
    urlcolor={blue},
    anchorcolor={black}
}

\makeatletter

\makeatletter

\begin{document}

\title{TRK Sum Rule for Interacting Photons}

	\author{Salvatore Savasta}
\affiliation{Dipartimento di Scienze Matematiche e Informatiche, Scienze Fisiche e  Scienze della Terra,
	Universit\`{a} di Messina, I-98166 Messina, Italy}
\author{Omar Di Stefano}
\email[corresponding author: ]{odistefano@unime.it}
\affiliation{Dipartimento di Scienze Matematiche e Informatiche, Scienze Fisiche e  Scienze della Terra,
	Universit\`{a} di Messina, I-98166 Messina, Italy}
\affiliation{Theoretical Quantum Physics Laboratory, RIKEN Cluster for Pioneering Research, Wako-shi, Saitama 351-0198, Japan}

\author{Franco Nori}
\affiliation{Theoretical Quantum Physics Laboratory, RIKEN Cluster for Pioneering Research, Wako-shi, Saitama 351-0198, Japan} \affiliation{Physics Department, The University
	of Michigan, Ann Arbor, Michigan 48109-1040, USA}




\begin{abstract}
The Thomas-Reiche-Kuhn sum rule is a fundamental consequence of the position-momentum commutation
relation for an atomic electron and it provides an important constraint on the transition matrix
elements for an atom.
Here we propose a TRK sum  rule for electromagnetic fields which is valid even in the presence of very strong light-matter interactions and/or optical nonlinearities.
While the standard TRK sum rule involves
dipole matrix moments calculated between  atomic energy levels (in the absence of interaction with the field), the sum rule here proposed involves expectation values of field operators calculated between general eigenstates of the interacting light-matter system.
This sum rule provides constraints and guidance for the analysis of strongly interacting light-matter systems and can  be used to test the validity
of approximate effective Hamiltonians often used in quantum optics. 
\end{abstract}

\maketitle

\section{Introduction}
\subsection{A Brief History of Sum Rules in Quantum Mechanics}
Since the beginning of quantum mechanics,  sum rules have proved to be very useful for understanding the general features of difficult problems. These relations, obtained by adding (sum) unknown terms, power tool for the study of physical processes \cite{Orlandini1991}. 
Historically, the first important sum rule is found in atomic physics and concerns the interaction of electromagnetism with atoms: the Thomas-Reiche-Kuhn (TRK) sum rule \cite{Thomas1925,Kuhn1925,Reiche1925}. It states that the sum of the squares of the dipole matrix
moments from any energy level, weighted by the corresponding energy differences, is a constant.
The TRK and analogous sum rules, like the Bethe sum rule \cite{Bethe1930}, play an especially important role in the interaction
between light and matter. They have widely been
applied to the problems of electron excitations in atoms,
molecules, and solids \cite{Wang1999}.

For an atomic electron,  the TRK sum rule is
a direct consequence, of the
canonical commutation relation between  position and
momentum. It is possible to view it as a necessary condition in order not to violate  this commutation relation \cite{Barnett1996}.
Among the many consequences of this sum rule,  it constrains the cross sections for absorption and
stimulated emission \cite{Merzbacher1970}. It has also been shown that useful sum rules can  be obtained for nonlinear optical susceptibilities \cite{Bassani1991,Scandolo1992,Scandolo1995}.
A modified TRK sum rule for the motion of the atomic center
of mass and a generalized TRK sum rule to include ions have
been also obtained \cite{Baxter1994}.  
Extensions of the TRK sum rule to  the
relativistic case have been studied (see, e.g., \cite{Levinger1957, Friar1975}). Important sum rules have also been developed in quantum chromodynamics (see, e.g., \cite{Nielsen2010}).

Such sum rules also play a relevant role in the analysis of interacting electron systems \cite{Pines1966, Giuliani2005}. Since they are a direct consequence of particle conservation in the system, their satisfaction is necessary to guarantee a gauge-invariant theory \cite{Pines1966, Giuliani2005} (see, e.g.,  Refs.~\cite{Andolina2019, Garziano2020} as two recent examples). 
In interacting electron systems, the longitudinal version of the TRK sum rule (known as $f$-sum rule) provides a very useful check on the
consistency of any approximate theory and can permit a direct calculation of collective mode frequencies in the long wavelength limit \cite{Pines1966}. A striking example of the relevance of sum rules in interacting electron systems is constituted by the apparent gauge invariance difficulty in superconductors (Meissner effect), originating by the violation of the $f$-sum rule of approximate models \cite{Anderson1958}.

Almost all the developed sum rules have been derived for the  degrees of freedom of particles. One exception is in Ref.~\cite{Barnett2012}, where optical sum rules have been derived  for polaritons propagating through a linear medium.

\subsection{Summary of our Main Results}
{\em Here we propose a TRK sum  rule for electromagnetic fields which is valid even in the presence of very strong light-matter interactions and/or optical nonlinearities} \cite{Kockum2018,Forn-Diaz2018}.
While the standard TRK sum rule involves
dipole matrix moments calculated between  atomic energy levels (in the absence of interaction with the field), the sum rule here proposed involves the expectation values of  the field coordinates or momenta calculated between general eigenstates of the interacting light-matter system (dressed light-matter states) and the corresponding eigenenergies of the interacting system. 

In this work, we also present a generalized atomic TRK sum rule for atoms strongly interacting with the electromagnetic field. This sum-rule has the same form of the standard TRK sum rule, but involves the energy eigenstates and eigenvalues of the interacting system.

The sum rules for interacting light-matter systems proposed here can  be useful to analyze general quantum  nonlinear optical effects (see, e.g., \cite{Peyronel2012,Chang2014,Guerreiro2014,Kockum2017a}) and many-body physics in photonic systems \cite{Carusotto2013}, like  analogous sum-rules for interacting electron systems, which  played a fundamental role for understanding the many-body physics of electron liquids \cite{Anderson1958, Pines1966, Giuliani2005}. The proposed sum rules, become particularly interesting in the non-perturbative regimes of light-matter interaction.

In the last years, several methods to control the  strength of the light-matter interaction have been developed, and the ultrastrong coupling (USC) between light and matter has transitioned
from theoretical proposals to experimental reality \cite{Kockum2018,Forn-Diaz2018}. In this new regime of quantum light-matter interaction,  beyond weak and strong coupling, the coupling strength becomes comparable to the transition frequencies in the system, or even higher [deep strong coupling (DSC)] \cite{DeLiberato2014, Garcia-Ripoll2015,Bayer2017,Yoshihara2017a}.
In the USC and DSC regimes, approximations widely employed in quantum optics break down \cite{Ridolfo2012}, allowing processes that do not conserve the number of excitations in the system (see, e.g., \cite{Niemczyk2010, Casanova2010,Garziano2016,Kockum2017a,Stassi2017}). The non-conservation of the excitation number gives rise to a wide variety of novel and unexpected physical phenomena in different hybrid quantum systems \cite{DeLiberato2007,Ashhab2010,Casanova2010,Carusotto2012,Auer2012,Garziano2013,Stassi2013,Garziano2014,Huang2014,Benenti2014,Garziano2015,Stassi2016,Jaako2016,DeLiberato2017,Cirio2017,Kockum2017,Albarr_n_Arriagada_2017,DiStefano2017,Macri2018a,Zheng2018,Felicetti2018,Macri2018}. 
As a consequence, all the system eigenstates, dressed by the interaction,  contain different numbers of excitations.
Much research on these systems has dealt with understanding
whether these excitations  are real or virtual,
how they can be probed or extracted, how they make
possible higher-order processes even at very low excitation densities, and how they affect the description of input and output for the system \cite{Kockum2018,Forn-Diaz2018}.

The eigenstates of these systems, including the ground state, can display a complex structure involving superposition of several eigenstates of the non-interacting subsystems \cite{Kockum2018,Forn-Diaz2018,Flick2018}, and can be difficult to calculate.  As a consequence, a number of approximation methods have been developed \cite{SanchezBurillo2019, Mordovina2019}. Moreover, the output field correlation functions, connected to measurements, depend on these eigenstates (see, e.g., Ref.~\cite{Stassi2016,DiStefano2017a}). 
Hence sum rules providing general guidance and constraints can be very useful to test the validity of the approximations.
The general sum rule proposed in this article
 can also be used to test the validity
of  effective Hamiltonians often used in quantum optics and cavity optomechanics \cite{Law1995,Macri2018,DiStefano2019prl}. 
In addition, this generalized TRK sum-rule applies to the broad emerging field of nonperturbative light-matter interactions, including several settings and subfields, as cavity and circuit-QED \cite{Kockum2018}, collective excitations in solids \cite{Kirton2019}, optomechanics \cite{Law1995}, photochemistry and QED chemistry \cite{Flick2017, Flick2018}. 

\section{Sum rule for interacting photons}
A key property used for the derivation of the TRK sum rule is that the commutator between the electron coordinate and the electronic Hamiltonian does not depend on the electronic potential, which is a function of the coordinate only, and hence it is universal. Considering for simplicity a single electron 1D system,  if $x$  is the electron coordinate and $\hat H_{\rm at} = \hat p^2 / 2m + V(x)$ is  the electronic Hamiltonian: $[x, \hat H_{\rm at}] $ $= [x,  \hat p^2 / 2m]$ $= i (\hbar/m) \hat p$.

In the Coulomb  gauge, the (transverse) vector potential ${\bf A}$ represents the field coordinate, while its conjugate momentum $\bf \Pi$ is proportional to the transverse electric field:
\be\label{momentum}
{\bf \Pi}({\bf x}, t)=-\varepsilon_0 \hat  {\bf  E}({\bf x}, t)= \varepsilon_0\dot{ {\hat {\bf A}}}({\bf x}, t) \, .
\ee
A general feature of the light-matter interaction Hamiltonians derived from the {\em minimal coupling replacement} (as for the Coulomb gauge) is that  the momenta of the matter system are coupled {\em only} to the field coordinate. We can express the   total light-matter quantum Hamiltonian as $\hat H  = \hat H_F + \hat H_M + \hat H_I$, where the first two terms on the  r.h.s. are the field and matter system free Hamiltonians, and the third describes the light-matter interaction.
Using \eqref{momentum} and the Heisenberg equation $i \hbar {\dot{ {\hat {\bf A}}}}= [{\hat {\bf A}}, \hat H]$, we obtain the relation
\be \label{m1}
i \hbar\,  {\bf \Pi} =
\varepsilon_0  [\hat {\bf A}, \hat H] = \varepsilon_0 [\hat {\bf A}, \hat H_F] \,  ,
\ee
where the second equality follows from  $ [\hat {\bf A}, \hat H_I] =0$, which holds, e.g.,  in the Coulomb gauge.
For simplicity, we consider the case of a quasi $1D$ electromagnetic resonator of length $L$, so that the expression for the electric-field operator can be simplified to $\hat  {\bf E} ({\bf r}, t) \to \tilde {\bf s} \hat  {E} ({x}, t)$, where $\tilde {\bf s} = {\bf y}/ |{\bf y}|$, where $x$ is the coordinate along the cavity axis, and $y$ a coordinate along an axis orthogonal to the cavity axis. The 
vector potential (as well as the electric field operator)
can be expanded in terms of photon creation and destruction operators as 
$$
\hat  { A} ({x}, t) =  \sum_m A_m (x) \hat a_m e^{-i\omega_m t} + {\rm h. c.}
$$ and
$$
\hat  { E} ({x}, t) =  \sum_m E_m (x) \hat a_m e^{-i\omega_m t} + {\rm h. c.}\,,,$$ where
$$A_m(x) =  [ {\hbar} /({2 \omega_m \varepsilon_0 S})]^{1/2} u_m (x)\,,$$ and  $$E_m(x) = i \omega_m  A_m(x)\,.$$ Here,
$S L$ is the resonator volume, the subscript $m$ labels a generic mode index with frequency $\omega_m$, and ${u}_m({ x})$ are the normal modes of the field chosen as real functions. For example, imposing the vanishing of the electric field at the two end walls at $x=\pm L/2$ of the cavity, $${u}_m({ x}) = 
(1/ \sqrt{L})\sin k_m (x + L/2)\,,$$ where $k_m = \pi m /L$.

Let us now consider the matrix elements of the operators in \eqref{m1} between two generic eigenstates $| \psi_i \rangle$ of the {\em total} Hamiltonian $\hat H$. We obtain
\be \label{R1}
{\bf \Pi}_{ij}=
i\varepsilon_0 \omega_{ij} {\bf A}_{ij}\, ,
\ee
where $\omega_{ji}=\omega_{j}-\omega_{i}$ and we used the notation $O_{ij}=\brakkets{\psi_i}{{\hat O}}{\psi_j}$. Here and in the following,  $j = 0$ indicates the system ground state, and  the energy levels are ordered  according to their energy: $j > i$ if $\omega_j > \omega_i$.
We now multiply both sides of \eqref{R1} by $u_m(x)$ and integrate over $x$. By defining $$\hat {\cal Q}^{(m)} = (\hat a_m + \hat a_m^\dag)/ \sqrt{2}\,,$$ and $$\hat {\cal P}^{(m)} = i(\hat a^\dag_m - \hat a_m)/ \sqrt{2}\,,$$ we obtain the corresponding relation for the individual modes:
\be\label{relationpq}
 \omega_m {\cal P}^{(m)}_{ij}    = i \omega_{ij} {\cal Q}^{(m)}_{ij}\, .
\ee
 It is worth noticing that, in the limit when the light-matter interaction vanishes, $| {\cal P}^{(m)}_{ij}|  =  |{\cal Q}^{(m)}_{ij}|$, and \eqref{relationpq} can easily be verified analytically . When the interaction becomes relevant, so that the system eigenstates differ from the harmonic spectrum for free fields, the ratio between the two quadratures can be very different from $1$ and can be determined by the only knowledge of the energy spectrum, independently on the  specific interacting system. Equation~(\ref{relationpq})  is the first result of this work. It shows that the ratio between the two field quadratures is uniquely determined by the energy spectrum. The two quadratures can display very different matrix elements when the interaction with the matter-system changes significantly the energy levels of the interacting systems, as it occurs in the USC and DSC regimes.

Let us now consider the commutator between the mode coordinate and its conjugate momentum:
\be \label{double}
i = \left [  \hat {\cal Q}^{(m)} ,  \hat {\cal P}^{(m)} \right]
 = 
  \frac{1}{ i \hbar \omega_m}  \left[ \hat {\cal Q}^{(m)} , \left[ \hat {\cal Q}^{(m)}, \hat H_F
 \right] \right] \, ,
\ee
where we used 
$$\omega_m \hat {\cal P}^{(m)}= \dot{\hat{{\cal Q}}}^{(m)}\, ,\: \text{and}\: \left[ \hat {\cal Q}^{(m)}, \hat H
\right]  = \left[ \hat {\cal Q}^{(m)}, \hat H_F
\right]\,.$$
 Developing the double commutator,
 considering its matrix elements between two generic eigenstates of the total Hamiltonian $\hat H$, and inserting the identity operators ($\hat I = \sum_k | \psi_k \rangle \langle \psi_k|$), we obtain the following relation
\be
\sum_k  \frac{ \omega_{k,i} + \omega_{k,j}}{\omega_m} 
{\cal Q}^{(m)}_{i,k}  {\cal Q}^{(m)}_{k,j}
= \delta_{i,j}\, ,
\ee
which reduces (choosing $j=i$) to the TRK sum rule for interacting fields:
\be \label{SDN}
2 \sum_k  \frac{ \omega_{k,i}}{\omega_m}  |{\cal Q}^{(m)}_{i,k}|^2 = 1\, .
\ee
By using \eqref{relationpq}, \eqref{SDN} can  be also expressed in terms of the momenta matrix elements: $$2 \omega_{m} \sum_k   |{\cal P}^{(m)}_{i,k}|^2/  \omega_{k,i} = 1.$$
Formally, it coincides with the TRK sum rule for atoms; however, in \eqref{SDN} the  matrix elements of the field-mode coordinate replace the atomic {\em electric-dipole} matrix elements. An important difference is that the atomic TRK sum rule \cite{Sakurai1994} considers atomic energy eigenstates, calculated in the absence of interaction with the field. On the contrary, this sum rule  is very general, since it holds in the presence of interactions with {\em arbitrary} matter systems, every time the interaction occurs via the field coordinate (e.g., Coulomb gauge). We also observe that \eqref{SDN} describes a collection of sum rules, one for each field mode $m$. Actually, following the same reasoning which led us to \eqref{SDN}, a generalized atomic TRK sum rule for atoms strongly interacting with the electromagnetic field [analogous to \eqref{SDN}] can be easily obtained, as shown in Sect.~\ref{TRKA}.

\section{Applications}

\subsection{Quantum Rabi model} 
\begin{figure}[htbp]
\centering
	\includegraphics[width= 1\linewidth]{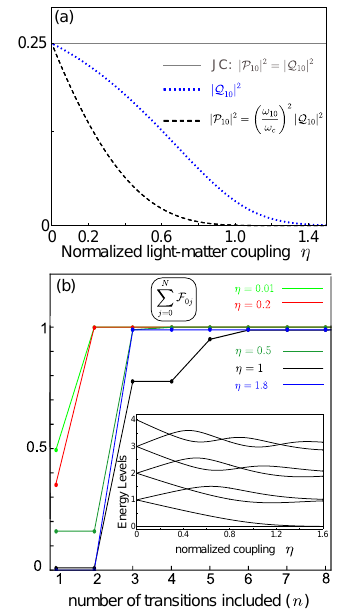}
	\caption{ (a)  $ {\cal P}$--${\cal Q}$ relation:   calculation of  $|{\cal P}_{1,0}|^2$ (proportional to the photodetection rate for cavity photons) (red continuous curve) and of  $|{\cal Q}_{1,0}|^2$ (dashed blue) versus the normalized coupling $\eta$. (b)  TRK sum rule for interacting fields: partial sums $\sum_{j= 0}^N {\cal F}_{0j}$ as function of the number $N$ of levels included for different normalized coupling rates $\eta$. Inset:  energy spectrum for the first energy levels $\omega_{k,0}$ versus the normalized coupling strength. 
	}
		\label{fig:Rabi}
\end{figure}

The quantum Rabi Hamiltonian, describes the
dipolar coupling between a two-level atom and a single mode of the quantized electromagnetic field. Recently, it has been  shown \cite{DiStefano2019}  that the correct (satisfying the gauge principle) quantum Rabi Hamiltonian  in the Coulomb gauge 
\bea\label{Hc}
\hat {H}_C = \hbar \omega_c \hat a^\dag \hat a &+& \frac{\hbar \omega_{0}}{2}  \left\{ \hat \sigma_z \cos{\left[ 2 \eta (\hat a + \hat a^\dag)\right]} \right. \nonumber \\
&+& \left. \hat \sigma_y \sin{\left[ 2 \eta (\hat a + \hat a^\dag)\right]} \right\}\,  ,
\eea
strongly differs from the standard model (see also  Refs~\cite{DeBernardis2018,Stokes2018, Settineri2019} for gauge issues in the USC regime). Here, $\omega_c$ is the resonance frequency of the cavity mode, $\omega_0$ is the transition frequency of a two-level atom,  $\hat a$ and $\hat a^\dag$ are the destruction and creation operators for the cavity field, while the qubit degrees of freedom are described by the Pauli operators $\hat \sigma_i$. 
The parameter $$\eta =  A_0 d/\hbar$$ ($A_0$ is the zero-point-fluctuation amplitude of the field potential and $d$ is the atomic dipole moment)  in \eqref{Hc} describes the normalized light-matter coupling  strength.
 When the normalized coupling strength is small  ($\eta \ll 1$), considering only first order contributions in $\eta$, the standard interaction term $\hbar \omega_0 \eta (\hat a + \hat a^\dag) \hat \sigma _y$ is recovered. If the system is prepared in its first excited state, the photodetection rate for cavity photons  is proportional to $|{\cal P}_{1,0}|^2$ (see Ref.~\cite{Settineri2019,DiStefano2017a}).
Figure~\ref{fig:Rabi}(a) displays this quantity (black dashed curve) as well as $|{\cal Q}_{1,0}|^2$ (dotted blue) versus the normalized coupling $\eta$, calculated after the numerical diagonalization of \eqref{Hc}. The two quantities are equal only at negligible coupling. When the coupling strength increases, the two quantities provide very different results. 
However, in agreement with \eqref{relationpq}, the numerically calculated $(\omega^2_{1,0}/ \omega^2_c)| {\cal Q}_{1,0}|^2$  coincides with $|{\cal P}_{1,0}|^2$ (black dashed curve). In contrast, the Jaynes Cummings (JC) model
\[
\hat H_{\rm JC}  = \hbar \omega_c \hat a^\dag \hat a + \hbar \omega_{0}/{2}  \hat \sigma_z + \hbar \eta \omega_c (\hat a \hat \sigma_+ + {\rm h.c.} )\, ,
\]
violates \eqref{relationpq} providing coupling-independent values  $|{\cal Q}_{1,0}|^2 = |{\cal P}_{1,0}|^2$ [the horizontal line in \figref{fig:Rabi}(a)].

These findings show that, using the wrong quadrature (${\cal Q}$ instead of ${\cal P}$) for the calculation of the photodetection rate for systems in the USC regime, can result into significantly wrong results. This is a direct consequence of \eqref{R1}.

In order to understand how the sum rule in \eqref{SDN} applies to the quantum Rabi model,   we calculate partial sums with an increasing number of  states. 
Specifically, we calculate $$\sum_{j= 1}^N {\cal F}_{0j}\,,\: \text{ where}\quad 
{\cal F}_{0j}= 2 (\omega_{j,0}/{\omega_c})  |{\cal Q}_{0,j}|^2\,.$$  Here and in the following, the eigenstates of the total Hamiltonian, obtained for a given  coupling strength $\eta$, are labelled so that $i > j$ for $\omega_i > \omega_j$.
Differently from the  JC, the quantum Rabi model does not conserve the excitation number. Therefore, expectation values  like $ {\cal Q}_{0,j}$ (and hence   $ {\cal F}_{0,j}$) can be different from zero also for $j > 2$. 
Figure~\ref{fig:Rabi}(b) displays such partial sums as a function of the number of levels included, obtained for different values of $\eta$.
For small values  ($\eta = 0.01$) only the two lowest excited levels contribute  to the sum with approximately equal weights, in good agreement with the  JC model. For $\eta= 0.2$ still only two transitions contribute to the sum rule; however the second transition provides a larger contribution to the sum.  For $\eta = 0.5$, the contribution of the lowest energy transition become smaller, while ${\cal F}_{02} = 0$, owing to the parity selection rule. Note that, at $\eta = \eta_{\rm cr} \simeq 0.44$ there is a crossing between the levels $2$ and $3$ [see inset in \figref{fig:Rabi}(b)], so that, for $\eta > \eta_{\rm cr}$, state $|2 \rangle$ has the same parity of state $|0 \rangle$. It is sufficient to include ${\cal F}_{03}$ to approximately satisfy the sum rule. For  $\eta = 1$,  ${\cal F}_{0,1} $  is very small and  ${\cal F}_{0,2} = 0$. In this case the sum rule is satisfied mainly with the contributions ${\cal F}_{0,j}$ with $3 \leq j \leq 6$. Finally, for very high values of the normalized coupling strength ($\eta = 1.8$) only one contribution (${\cal F}_{0,3}$) becomes relevant. 
 This effect is due to the light-matter decoupling \cite{Settineri2019} which occurs at very high values of $\eta$, where the system ground state $|0 \rangle$  is well approximated by $|g,0 \rangle$  [the first entry in the ket labels the photon number, the second labels the qubit state: ground ($g$) or excited ($e$)], then  $|1 \rangle  \simeq |e,0 \rangle$, $|2 \rangle  \simeq |e,1 \rangle$, $|3 \rangle  \simeq |g,1 \rangle$, and so on: the higher energy levels are of the kind $\simeq |g(e),n>1 \rangle$. This explains why for $\eta=1.8$ the only significant contribution  to the sum is ${\cal F}_{0,3}$. These behaviours of the partial sums and of the terms ${\cal F}_{i,j}$ are closely connected to accessible  experimental features, as explicitly shown in the example below.

\subsection{Nonlinear electromagnetic resonator}\label{sec:3} 
\begin{figure}[htbp]
\centering
\includegraphics[width= 1.\linewidth]{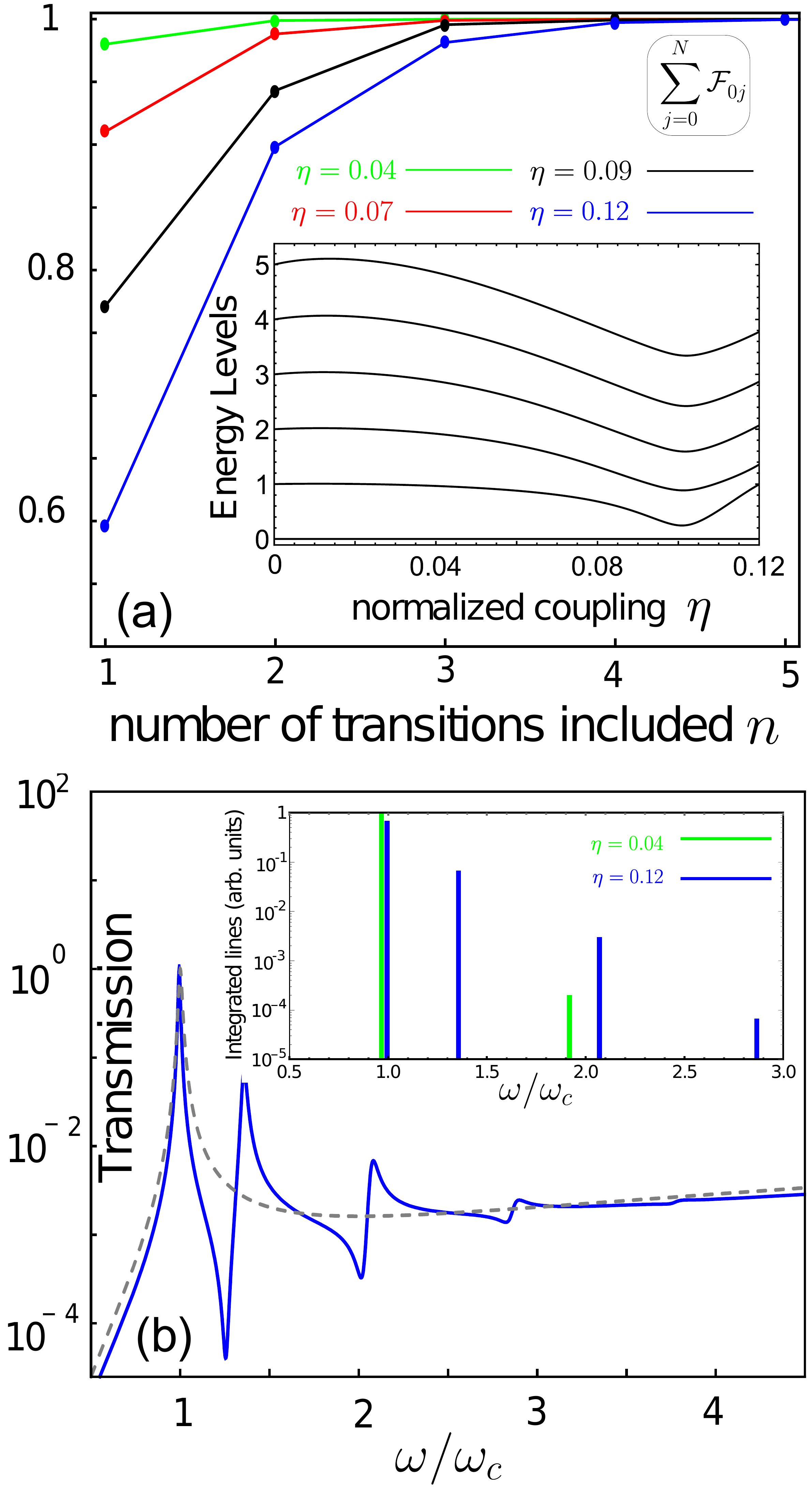}
	\caption{(a)  TRK sum rule for a single-mode nonlinear system: partial sums $\sum_{j= 0}^N {\cal F}_{0j}$ versus the number ($N$) of levels included for different normalized coupling strengths $\eta$. Inset: anharmonic energy spectrum $\omega_{k,0}$ versus $\eta$.
	(b)  Transmission spectrum $T(\omega)$ for a two-port nonlinear resonator for $\eta = 0.12$. The inset shows  the integrated lines for two values of $\eta$. }
		\label{fig:NLO}
\end{figure}
As a further test, we analyze a single-mode nonlinear optical system described by the following effective Hamiltonian
\be\label{Hnl}
\hat H= \hbar \omega_c \hat a^\dag \hat a
+ \eta \hbar \omega_c \left(\hat a +\hat a^\dag\right)^3
+ \frac{\eta}{10}  \hbar\omega_c \left( \hat a + \hat a^\dag\right)^4\,  .
\ee
Here $\hat H_F = \hbar \omega_c \hat a^\dag \hat a$, while the nonlinear terms are assumed to arise from the dispersive interaction with some material system \cite{Jacobs2009}. Note that the nonlinear terms in \eqref{Hnl} commutes with the field coordinate $\hat {\cal Q} = (\hat a + \hat a^\dag)/\sqrt{2}$, hence Eqs.~(\ref{relationpq}) and (\ref{SDN}) holds. In contrast, the presence of a standard self-Kerr term  $\propto \hat a^{\dag\, 2} \hat a^2$ (see, e.g., \cite{Ferretti2012}) would violate them.
The inset in \figref{fig:NLO} shows the anharmonic energy spectrum $\omega_{k,0}$ as a function of $\eta$. 
Figure~(\ref{fig:NLO}) displays the partial sums $\sum_{j= 1}^N {\cal F}_{0j}$ as versus the number of included levels, calculated for different values of $\eta$. Increasing the anharmonicity coefficient $\eta$, the number of contributions in the sum increases at the expense of the contribution ${\cal F}_{01}$ of the lowest energy transition.
This behaviour is closely connected with accessible  experimental features which can be observed, e.g.,  in linear transmission spectra. For a two-port (equally coupled to the external modes) nonlinear resonator, the transmission spectrum (see Supplement \ref{SM} for supporting content) can be written as
\be\label{T}
T(\omega) = \omega^2 \left|\sum_k 
\frac{\Gamma_{k,0}/\omega_{k,0}}{\omega_{k,0}- \omega - i \Gamma_k }
\right|^2\, ,
\ee
where the radiative decay rates are $$\Gamma_{k,j} = 2 \pi g^2(\omega_{k,j})\, |{\cal Q}_{k,j}|^2\,, \quad \Gamma_k = \sum_{j<k} \Gamma_{k,j}\,,$$ and we assumed an ohmic coupling with the external modes ($g^2(\omega) \propto \omega$).
When the anharmonicity is switched off ($\eta = 0$), $\Gamma_{k,0} \propto {\cal F}_{0k} = 0$ for $k \neq 1$, and  the transmission spectrum presents a single peak at $\omega = \omega_c$ [dashed curve in \figref{fig:NLO}(b)]. 
When $\eta \neq 0$,  $\Gamma_{k,0} \propto {\cal F}_{0k} \neq 0$, and the transmission spectrum in \figref{fig:NLO}   evolves accordingly (the blue-continuous curve show the spectrum calculated for $\eta = 0.12$). By integrating the individual spectral lines in \eqref{T}, we obtain for each line a contribution $\simeq \pi \Gamma^2_{k,0}/\Gamma_k$, which is approximately proportional to ${\cal F}_{0k}$ in the sum (notice that $\Gamma_k \sim k \Gamma_1$). The inset in \figref{fig:NLO} shows the integrated lines for two values of $\eta$.

\subsection{Frequency conversion in ultrastrong cavity QED}

The relations in \eqref{relationpq} and \eqref{SDN} are very general. So far we applied them to single-mode fields, however they are also  valid in the presence of   (even interacting) multi-mode fields (see, e.g., \cite{Malekakhlagh2016,SanchezMunoz2018}).
Here we analyze the TRK sum rule for interacting photons in a three-component system constituted by two single-mode  resonators  ultrastrongly coupled to a single superconducting flux qubit. This coupling can induce an effective interaction between the fields of the two resonators. Using suitable parameters for the three components, the system provides a method for frequency conversion of photons which is both versatile and deterministic. It has been shown that it can be used to realize both single and multiphoton frequency conversion processes \cite{Kockum2017}. The system Hamiltonian is  
\bea\label{HA1}
&&\hat H=\hbar \omega_a \hat a^\dag \hat a +\hbar \omega_b  \hat b^\dag \hat b
+ \frac{\hbar \omega_0}{2} \hat \sigma_z 
+  \\&& \hbar \left[g_a \left(\hat a +\hat a^\dag\right)+ g_b \left(\hat b +\hat b^\dag\right)\right]\left[\cos(\theta)\hat \sigma_x +\sin(\theta)\hat \sigma_z \right]
\,  ,\nonumber
\eea
where ($\hat a$, $\omega_a$, $g_a$) and  ($\hat b$, $\omega_b$, $g_b$) describe the photon operator, the frequency mode, and the coupling with the qubit for the two resonators. The angle $\theta$ encodes the qubit flux offset which determines parity symmetry breaking. A zero flux offset implies $\theta = 0$.
\begin{figure}[htbp]
\centering
	\includegraphics[width=1. \linewidth]{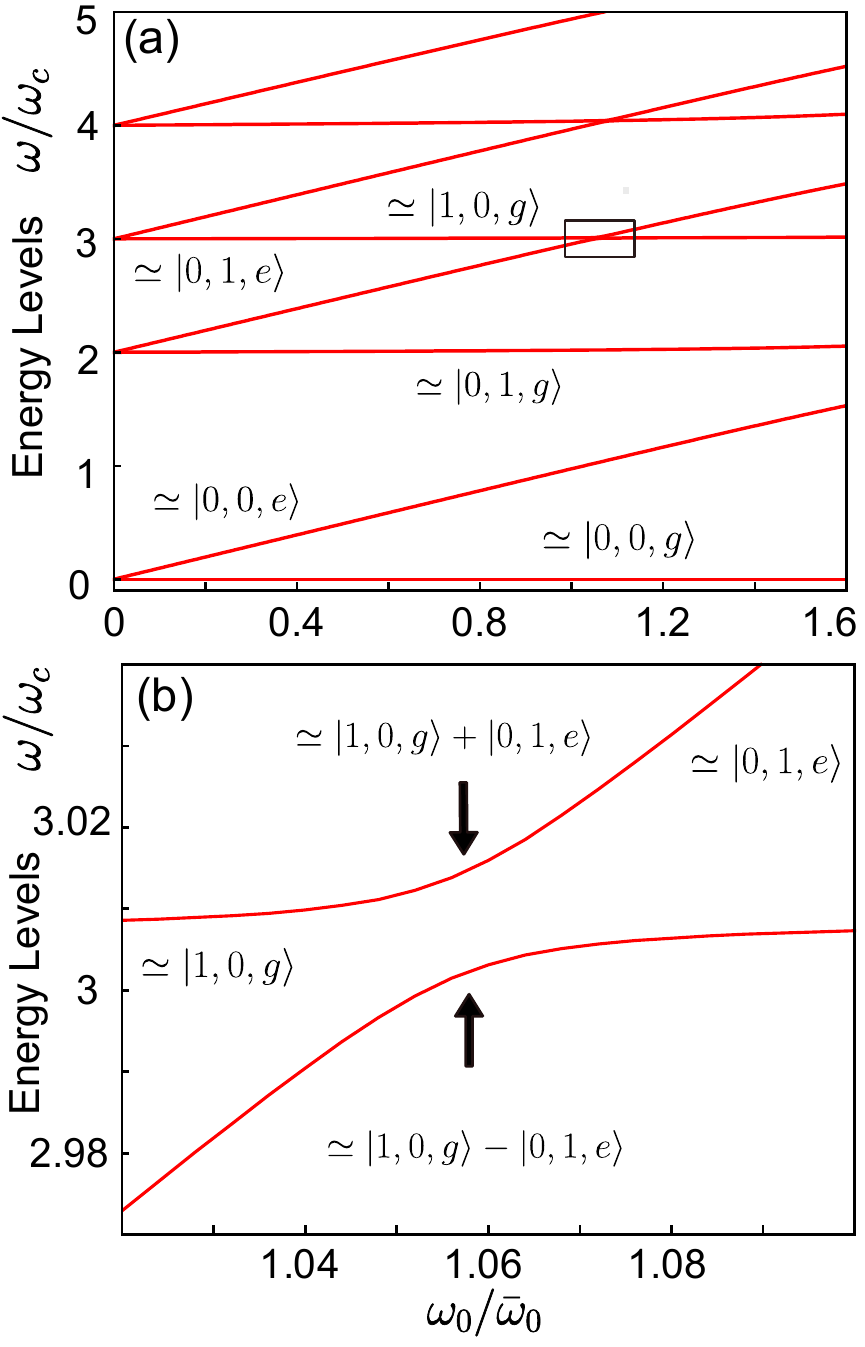}
	\caption{{Energy spectum obtained from the numerical diagonalization of \eqref{HA1}}. (a) Lowest normalized energy levels  versus the  qubit frequency.
		(b) Enlarged view of the spectrum inside the rectangle  in (a) showing the presence of an avoided level crossing. 
		Parameters are given in the text.}
	\label{fig:3}
\end{figure}
Figure~\ref{fig:3}(a) displays the lowest normalized energy levels  $(\omega - \omega_g)/\bar \omega_0$ (we indicated with $\hbar \omega_g$ the ground state energy) versus the  qubit frequency $\omega_0 / \bar \omega_0$ obtained diagonalizing numerically the Hamiltonian in \eqref{HA1}. We used the parameters $\omega_a= 3 \bar \omega_0$, $\omega_b= 2 \bar \omega_0$, $\theta=\pi/6$, $g_a= g_b=0.2\, \bar \omega_0$, where $\bar \omega_0$ is a  reference point for the qubit frequency. Notice that the two resonators are set in order that their resonance frequencies  satisfy the relationship $\omega_a = \omega_b + \bar \omega_0$.
The first excited level is a line with slope $\simeq 1$, corresponding to the approximate eigenstate $| \psi_1 \rangle \simeq |0,0,e \rangle$, where the first two entries in the ket indicate the number of photons in resonator $a$ and $b$ respectively, while the third entry indicates the qubit state. The second excited level is a horizontal line corresponding to 
the eigenstate $| \psi_2 \rangle \simeq |0,1,g \rangle$, the next two lines on the left of the small rectangle in \figref{fig:3}(a) (for values of $\omega_0 / \bar \omega_0$ before the apparent crossing),  correspond to the states $| \psi_3 \rangle \simeq |0,1,e \rangle$ and $| \psi_4 \rangle \simeq |1,0,g \rangle$. The apparent crossing in the rectangle is actually an avoided level crossing, as can be inferred from the enlarged view in \figref{fig:3}(b). It arises from the hybridization of the states  $|0,1,e \rangle$ and $|1,0,g \rangle$ induced by the counter-rotating terms in the system Hamiltonian. The resulting eigenstates can be approximately written as
\bea\label{34}
&&| \psi_{3} \rangle  \simeq
\cos \theta\,|0,1,e \rangle
-  \sin \theta \,|1,0,g \rangle \nonumber \\
&& | \psi_{4} \rangle \simeq
\sin \theta\,|0,1,e \rangle
+  \cos \theta \,|1,0,g \rangle\, .
\eea
The mixing is maximum when the level splitting is minimum (at $\omega_0 / \bar \omega_0 \simeq 1.056$). In this case $\theta = \pi /4$.

It has been shown \cite{Kockum2017} that this effective coupling can be used to transfer a quantum state constituted by an arbitrary superposition of zero  and one photon in one resonator (e.g., $a$), to a quantum state corresponding to the same superposition in the resonator at frequency $\omega_b$.

\begin{figure}
	\includegraphics[width= 1\linewidth]{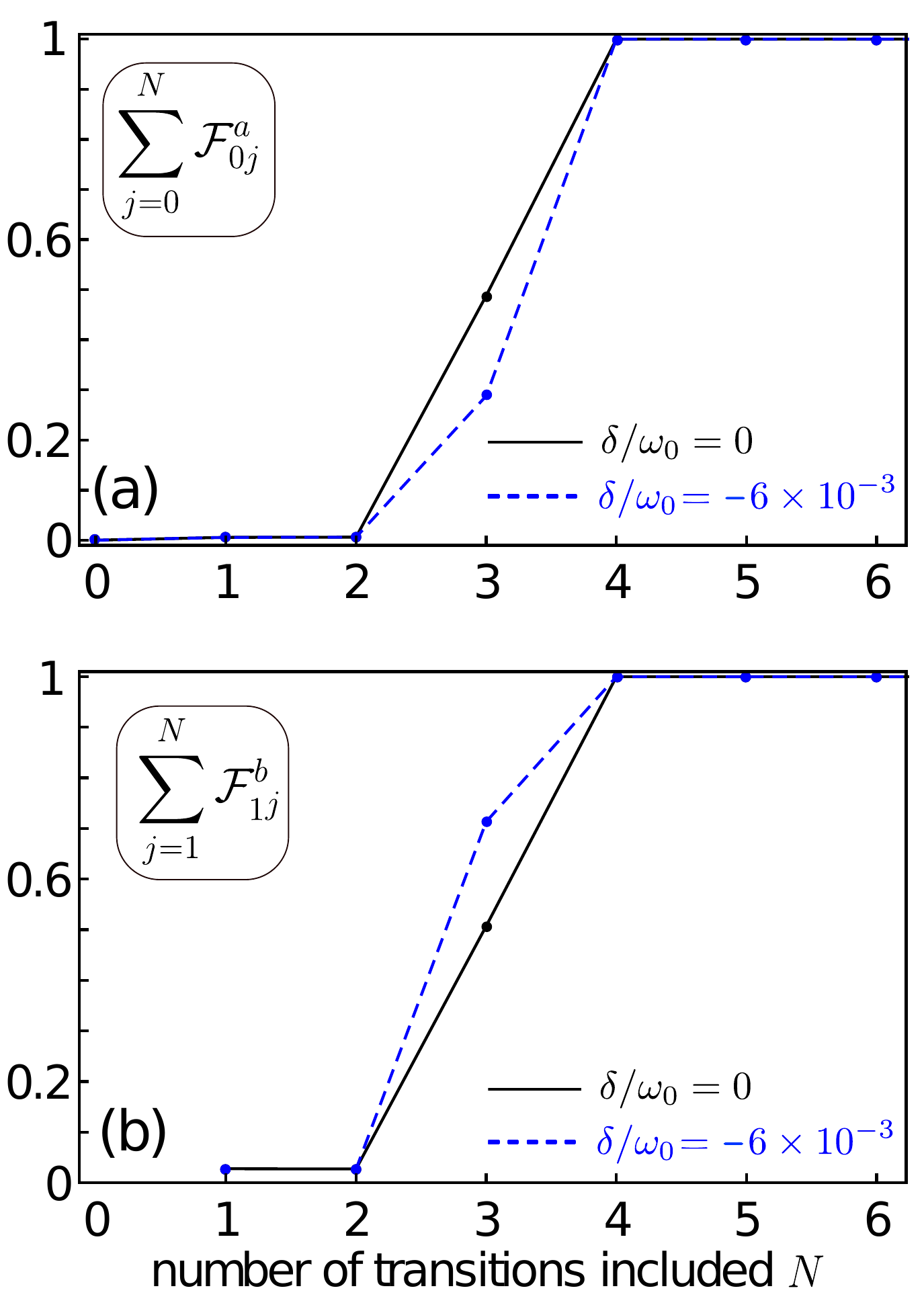}
	\caption{{TRK sum rule for interacting photons in the three-component system described by the Hamiltonian in \eqref{HA1}}. 
		(a) Partial sum rules $\sum_{j= 1}^N {\cal F}^a_{0j}$ relative to the first resonator and (b) $\sum_{j= 1}^N {\cal F}^b_{1j}$ relative to the second resonator, both for different values of levels $N$. The black segmented line describes the zero detuning case $\delta=0$, while the dashed blue segmented lines refer to the case $\delta = (\omega_0 - \bar \omega_0)/ \bar \omega_0 = - 6 \times 10^{-3}$. 
		Parameters are given in the text.}
	\label{fig:NLO2}
\end{figure}

This system represents an interesting example of two interacting optical modes (with the interaction mediated by a qubit). 
In order to understand how the sum rule in \eqref{SDN} applies to such a system, we investigate its convergence, calculating  partial sum rules for 
the two modes. Figure~\ref{fig:NLO2} shows $\sum_{j= 0}^N {\cal F}^a_{0j}$ (a) and $\sum_{j= 1}^N {\cal F}^b_{1j}$ (b) for different values of $N$. The black line describes the zero detuning case, while the dashed blue line the case $\delta = (\omega_0 - \bar \omega_0)/ \bar \omega_0 = - 6 \times 10^{-3}$. The results in \figref{fig:NLO2}(a) can be understood observing that
$$ { \cal F}^a_{0j} \propto |\langle 0 | \hat a + \hat a^\dagger| j \rangle|^2\,.$$
Since
$$| 0 \rangle \simeq |0,0,g \rangle\,, | 1 \rangle \simeq |0,0,e \rangle\,, | 2 \rangle \simeq |0,1,g \rangle\,, | 3 \rangle\,, \text{and}\, | 4 \rangle$$ are provided in \eqref{34}, it is easy to obtain 
$${\cal F}^a_{01} \simeq {\cal F}^a_{02} \simeq 0\,,  {\cal F}^a_{03} \propto \sin^2 \theta\,, \text{and}\, {\cal F}^a_{04} \propto \cos^2 \theta\,,$$ in agreement with the results in \figref{fig:NLO2}(a). Notice that for $\delta= 0 $, it results in $\theta = \pi/4$, and hence ${\cal F}^a_{03} \simeq {\cal F}^a_{04}$. A similar analysis can be carried out for the results in \figref{fig:NLO2}(b).

\section{TRK sum rule for atoms interacting with photons}\label{TRKA}
The standard atomic TRK sum rule \cite{Sakurai1994} considers atomic energy eigenstates, calculated in the absence of interaction with the transverse electromagnetic field. A recent interesting example of  descriptions including the electron-electron interaction can be found in Ref.~\cite{Andolina2019}.

Following the same reasoning which led us to \eqref{SDN}, a generalized atomic TRK sum rule for atoms strongly interacting with the electromagnetic field [analogous to \eqref{SDN}] can be easily obtained, starting from the dipole gauge. In this gauge (see, e.g., Ref.~\cite{DiStefano2019}), the light-matter interaction term does not depend on   the particle momentum, and the same steps used to obtain  \eqref{SDN} can thus be followed. The resulting atomic generalized TRK sum rule, formally coincides with the standard one, with the only difference that all the expectation values are calculated using the eigenstates of the {\em total} light-matter system. For example, we consider a system described by a single effective particle with mass $m$ and charge $q$ displaying a dipolar interaction with a single mode resonator:
\be
\hat H_D = \frac{1}{2 m} \hat p^2 + V(x) + \frac{q^2 \omega_c A^2_0}{\hbar} x^2 + i q \omega_c A_0 x (\hat a^\dag - \hat a)\,  ,
\ee
where $A_0$ is the zero-point-fluctuation amplitude of the field potential.
The following commutation relation holds:  $[x, \hat H_{D}] = [x,  \hat p^2 / 2m]= i (\hbar/m) \hat p$. From it, following the same steps  used to obtain  \eqref{SDN} or to obtain the standard atomic TRK sum rule, we obtain the TRK sum rule for a dipole interacting with the electromagnetic field:
\be\label{SDNa}
2 m  \sum_k \omega_{k,j} |x_{k,j}|^2 = 1\, ,
\ee 
where $x_{k,j}  \equiv \langle i |x| j \rangle$ is the expectation value of the position operator between two dressed states. Following the same reasoning, it can also be shown that also the $f$-sum rule \cite{Pines1966} (the longitudinal analog of the TRK sum rule) for an electron system strongly interacting with a quantized electromagnetic field can be obtained. These sum rules can find useful applications in the study of correlated electron systems strongly interacting with photons (see, e.g., \cite{Knuppel2019}).

\section{Discussion}

 The TRK sum rule for interacting photons proposed here can  be useful for investigating general quantum  nonlinear optical effects  and many-body physics in photonic systems (see, e.g., \cite{Peyronel2012, Chang2014,Guerreiro2014,Kockum2017a, Carusotto2013}), like  the corresponding sum-rules for interacting electron systems, which  played a fundamental role for understanding the many-body physics of interacting electron systems \cite{Anderson1958, Pines1966, Giuliani2005}.
 
 We provided a few examples showing how the light-matter interaction can change significantly the number of excited photonic states exhausting the sum rule. Using the sum rule, one can prove without explicit calculations that other excited states have negligible oscillator strength.

 The relations in \eqref{relationpq} and \eqref{SDN} are very general. They are also  valid in  systems including several dipoles (see, e.g., \cite{DeBernardis2018a, Bin2019}) and modes (see, e.g., \cite{SanchezMunoz2018}).
These relations provide a very useful check on the
consistency of approximate models in quantum optics.
Approximate Hamiltonians and effective models can violate one of them. Such a violation indicates that the model may miss some relevant physics \cite{Pines1966}. For example, we have shown that the JC model, a widespread description for the
dipolar coupling between a two-level atom and a quantized electromagnetic field, violates the relation \eqref{relationpq}. An additional example of a model violating this relation is provided by the well-known and widely employed cavity optomechanical interaction Hamiltonian  $\hbar g \hat a^\dag \hat a (\hat b + \hat b^\dag)$ (here $\hat b$ is the destruction operator for the mechanical oscillator) \cite{Aspelmeyer2014}. On the contrary,   the interaction Hamiltonian obtained by a microscopic model \cite{Law1995} 
$\hbar g (\hat a^\dag + \hat a)^2 (\hat b + \hat b^\dag)$, 
satisfies both of these relations [Eq.s~(\ref{SDN}), (\ref{SDNa})]. It turns out that such interaction Hamiltonian, in addition to the standard optomechanical effects,  also describes the dynamical Casimir effect \cite{Macri2018,DiStefano2019prl}.

An interesting feature of the relations proposed here is that they hold in the presence of light-matter interactions of arbitrary strength.
Moreover, the obtained sum rule can be useful for the analysis of strongly interacting light-matter systems, especially when exact eigenstates are not available.
These   relations in \eqref{relationpq} and \eqref{SDN} can provide constraints and a guidance in the development of effective Hamiltonians in quantum optics and cavity optomechanics.

Following the same reasoning leading to \eqref{SDN},  we also proposed a generalized TRK sum rule for the matter component involving transitions between the {\em total} light-matter  energy eigenstates [\eqref{SDNa}], describing particle conservation in the presence of arbitrary light-matter interactions.

\appendix
\label{SM}

\section{Linear response theory and transmission of a nonlinear optical system}
\label{a}
This section provides a derivation of the transmission coefficient of a nonlinear optical system based on the dressed master equation approach \cite{Beaudoin2011, Settineri2018}.

The dressed master equation in the Schr\"odinger picture  can be written as \cite{Beaudoin2011, Settineri2018},
\be\label{rhodot}
\dot{\hat \rho}(t) = -i\comm{\hat H_S}{\hat \rho(t)} + \mathcal{L} \hat \rho(t) \, ,
\ee
where $ \rho(t)$ is the density matrix operator for the nonlinear optical system,
\be
\hat H_S=\sum_k \omega_k |k\rangle \langle k|\, ,
\ee
is the system Hamiltonian expressed in the dressed basis, constituted by the  energy eigenstates of the nonlinear system.
Dissipation is described by the Lindbladian superoperator defined by
\bea
\mathcal{L} \hat \rho(t) =\sum\limits_{i}\sum\limits_{j, k < j} \left\{  \Gamma_{jk}^{(i)}\,  n (\omega_{jk}, T_i) 
 \lind{|j\rangle \langle k|} \hat \rho(t) + \right.\nonumber \\ \left. \Gamma_{jk}^{(i)} 
 \left[ 1 + n (\omega_{jk}, T_i) \right] \lind{|k\rangle \langle j|} \hat \rho(t) \right\}\, .
\label{lind1}
\eea
This equation  includes the thermal populations  
\be
n (\Delta_{jk}, T_i) = \left[\exp{\{\omega_{jk} /k_B T_i\}} - 1 \right]^{-1}\,,
\ee
and the the damping rates 
\be
  \Gamma_{jk}^{(i)} = 2 \pi g^2_i (\omega_{jk}) \abssq{X_{jk}} \, .
\ee
Here, $i=\{L, R\}$ indicates the input-output ports, $g (\omega)$ is  the system-reservoir coupling strength, $\hat X$ is the system operator interacting with the external modes, and
\be
\lind{\hat O} \hat \rho = \frac{1}{2} \left( 2 \hat O \hat \rho \hat O^\dag - \hat \rho \hat O^\dag \hat O - \hat O^\dag \hat O \hat \rho \right) \, .
\ee
At $T=0$, being $n (\Delta_{jk}, T_i) =0$, we obtain
\be
\mathcal{L} \hat \rho\:\underset {T=0}{=\joinrel=}\: \mathcal{L}_{0} \hat \rho\:=\: \sum_{i}\sum_{j, k < j} \left\{  \Gamma^{jk}_i  \lind{|k\rangle \langle j|} \hat \rho \right\} \, .
\label{lind2}
\ee

We also consider a coherent drive  entering from the left port, described by the following interaction Hamiltonian
\be\label{Hd}
\hat H_d(t) = i\hat X\int\!\!d\omega g_{\rm L}(\omega)[ e^{-i \omega t} \beta_L(\omega) - e^{i \omega t} \beta^*_L(\omega)]\, ,
\ee
where $\hat X$ is the system operator interacting with the external modes, and $$\beta_L(\omega) = \langle \hat b_L (\omega) \rangle$$ is a c-number corresponding to the mean value of the external (left) field operators, assumed to be in a coherent state. We will also assume $$\hat X=\hat {\cal Q} = (\hat a +\hat a^\dag) / \sqrt{2}\,,$$ where $\hat a$ is the photon destruction operator for a single-mode electromagnetic resonator.
The master equation~(\ref{rhodot}) becomes
\be\label{rhop}
\dot{\hat \rho}(t) = - i \comm{\hat H_S+\hat H_d(t)}{\hat \rho(t)} + \mathcal{L}_0 \hat \rho(t) \, .
\ee 
We assume that  the  light field from the left port is coherent with driving frequency $\omega$: $$\expec{\hat b_\omega}= \beta_{\rm L}(\omega)\exp{[-i\omega t]}\,.$$
Retaining only the terms depending linearly from the input field and using \eqsRef{rhodot,lind2,Hd}, assuming
$$\rho_{n0}(t)=\rho_{n0} \exp{[-i\omega t]}$$ (i.e., oscillating resonantly with the driving field), and using the rotating wave approximation, 
 we obtain (to first order in the field)
\be\label{rhon0}
\rho^{(1)}_{n0}= \frac{i g_{\rm L}(\omega) \beta_{\rm L}(\omega)X_{n0} }{(\omega-\omega_{n0})+i \sum_i\sum_{k<n}\Gamma^{n,k}_i }\, ,
\ee
where, being $T=0$, only the ground state is populated  in the absence of interaction ($\rho^{(0)}_{00} = 1$).
In order to calculate the transmitted signal that can be experimentally detected,
we consider a system constituted by an $LC$-oscillator coupled to  a transmission line 
and use the input-output relations \cite{Settineri2019} for the positive frequency component of the output (input) vector potential operator
defined as
\be\label{Pinout}
\hat \phi^{+}_{{\rm out}(\rm in)}(t)= \Lambda
\int_0^{\infty} \frac{d\omega}{\sqrt{\omega}}\,  \, \hat b_\omega^{\rm out (\rm in)}(t)\,  ,
\ee
where, for the sake of simplicity, we disregarded the spatial dependence, and $\Lambda = \sqrt{\hbar Z_0/4 \pi}$, with $Z_0$  the impedance of the in-out transmission line(s).
In addition, we consider two distinct ports for the input ($L$) and the output ($R$) [for simplicity we assume $g_{\rm L}(\omega)=g_{\rm R}(\omega)=g(\omega)$] and  we have
for the output voltage operator \cite{Settineri2019} $\hat V^{(\rm R)+ }_{\rm out}(t) = \dot{\hat \phi}^{(\rm R)+ }_{\rm out}(t)$:
\begin{equation}
\label{inpoutV}
\hat V^{(\rm R)+ }_{\rm out}(t) =  -2\pi \Lambda \sum_{ j} \frac{g(\omega_{j0})}{\sqrt{\omega_ {j0}}}  X_{0j}  \dot{\hat P}_{0j}(t)\,,
\end{equation}
which can be expressed as
\be\label{v2}
\hat V^{(\rm R)+ }_{\rm out}(t)=- K\,   \hat V^{+} (t)\, ,
\ee
where
\be\label{vL}
\hat V^+ = \Phi_{\rm zpf} \sum_ {j}  X_{0j} \dot{\hat P}_{0j}(t)\, .
\ee
Assuming $g(\omega)=G \sqrt{\omega}$, the constants $K$ and  $\Phi_{\rm zpf}$ satisfy the relation 
\be\label{K}
\frac{K  \Phi_{\rm zpf}}{\Lambda}= 2\pi G\, .
\ee
Using \eqref{Pinout}, we have for the mean value of the input sent through  the port ($L$) 
\be\label{phiin}
\expec{\hat V^{(L)+ }_{\rm in}(t)}
=\expec{ \dot{\hat \phi}^{(L) + }_{\rm in}(t)}
= -i \Lambda \sqrt{\omega}\,  \, \beta_{\rm L}({\omega})\,  ,
\ee
where we assumed a coherent drive input at frequency $\omega$: $$\expec{\hat b_{\omega'}^{\rm L}(t)}= \beta_{\rm L}({\omega})\delta(\omega'-\omega)\,.$$
Considering the linear response only, the projection operator oscillates at the frequency $\omega$ of the drive, $$\dot{\hat P}_{0j}(t)=-i\omega {\hat P}_{0j}(t)\,,$$  using \eqsRef{v2,vL}, the mean value for the output is 
\be\label{meanout}
\expec{\hat V^{(\rm R)+ }_{\rm out}(t)}=i K \Phi_{\rm zpf}\, \omega \sum_ {j}  X_{0j} \rho_{j0}(t)\, ,
\ee
where $\hat \rho$ is the density matrix and we used the  relation $$\expec{{\hat P}_{0j}(t)}=\rho_{j0}(t)\,.$$
Using \eqsRef{K,meanout,phiin}, we can calculate the transmission coefficient $T(\omega)$ due to the signal detected from  the port ($R$) when a driving field is sent through the port ($L$) as
\be\label{f}
T(\omega)=\!\!\abs{\frac{\expec{\hat V^{(\rm R)+}_{\rm out}(t)}}{\expec{\hat V^{(\rm L)+}_{\rm in}(t)}}}^2\!\!\!=\!\omega^2\abs{\sum_j\frac{\Gamma_{j0} /\omega_{j0}}{(\omega-\omega_{j0})+i\sum\limits_i\sum\limits_{k<n}\Gamma^{nk}_j}}^2\!\!\!\!,
\ee 
where $\Gamma_{j0}=2\pi \abs{g(\omega_{j0})}^2\abs{X_{j0}}^2$.
Recalling that we assumed $\hat X = \hat {\cal Q}$, \eqref{f} corresponds to \eqref{T}.

\vspace{1 cm}

{\bf Acknowledgments}

F.N. is supported in part by: NTT Research,
Army Research Office (ARO) (Grant No. W911NF-18-1-0358),
Japan Science and Technology Agency (JST)
(via 
the CREST Grant No. JPMJCR1676),
Japan Society for the Promotion of Science (JSPS) (via the KAKENHI Grant No. JP20H00134, and the grant JSPS-RFBR Grant No. JPJSBP120194828), and
the Grant No. FQXi-IAF19-06 from the Foundational Questions Institute Fund (FQXi),
a donor advised fund of the Silicon Valley Community Foundation.

S.S. acknowledges the Army Research Office (ARO)
(Grant No. W911NF1910065).



\bibliographystyle{apsrev4-2}
\bibliography{refMEnew2}

\begin{thebibliography}{81}%
\makeatletter
\providecommand \@ifxundefined [1]{%
 \@ifx{#1\undefined}
}%
\providecommand \@ifnum [1]{%
 \ifnum #1\expandafter \@firstoftwo
 \else \expandafter \@secondoftwo
 \fi
}%
\providecommand \@ifx [1]{%
 \ifx #1\expandafter \@firstoftwo
 \else \expandafter \@secondoftwo
 \fi
}%
\providecommand \natexlab [1]{#1}%
\providecommand \enquote  [1]{``#1''}%
\providecommand \bibnamefont  [1]{#1}%
\providecommand \bibfnamefont [1]{#1}%
\providecommand \citenamefont [1]{#1}%
\providecommand \href@noop [0]{\@secondoftwo}%
\providecommand \href [0]{\begingroup \@sanitize@url \@href}%
\providecommand \@href[1]{\@@startlink{#1}\@@href}%
\providecommand \@@href[1]{\endgroup#1\@@endlink}%
\providecommand \@sanitize@url [0]{\catcode `\\12\catcode `\$12\catcode
  `\&12\catcode `\#12\catcode `\^12\catcode `\_12\catcode `\%12\relax}%
\providecommand \@@startlink[1]{}%
\providecommand \@@endlink[0]{}%
\providecommand \url  [0]{\begingroup\@sanitize@url \@url }%
\providecommand \@url [1]{\endgroup\@href {#1}{\urlprefix }}%
\providecommand \urlprefix  [0]{URL }%
\providecommand \Eprint [0]{\href }%
\providecommand \doibase [0]{https://doi.org/}%
\providecommand \selectlanguage [0]{\@gobble}%
\providecommand \bibinfo  [0]{\@secondoftwo}%
\providecommand \bibfield  [0]{\@secondoftwo}%
\providecommand \translation [1]{[#1]}%
\providecommand \BibitemOpen [0]{}%
\providecommand \bibitemStop [0]{}%
\providecommand \bibitemNoStop [0]{.\EOS\space}%
\providecommand \EOS [0]{\spacefactor3000\relax}%
\providecommand \BibitemShut  [1]{\csname bibitem#1\endcsname}%
\let\auto@bib@innerbib\@empty
\bibitem [{\citenamefont {Orlandini}\ and\ \citenamefont
  {Traini}(1991)}]{Orlandini1991}%
  \BibitemOpen
  \bibfield  {author} {\bibinfo {author} {\bibfnamefont {G.}~\bibnamefont
  {Orlandini}}\ and\ \bibinfo {author} {\bibfnamefont {M.}~\bibnamefont
  {Traini}},\ }\href
  {https://iopscience.iop.org/article/10.1088/0034-4885/54/2/002} {\bibfield
  {journal} {\bibinfo  {journal} {Rep. Prog. Phys.}\ }\textbf {\bibinfo
  {volume} {54}},\ \bibinfo {pages} {257} (\bibinfo {year} {1991})}\BibitemShut
  {NoStop}%
\bibitem [{\citenamefont {Thomas}(1925)}]{Thomas1925}%
  \BibitemOpen
  \bibfield  {author} {\bibinfo {author} {\bibfnamefont {W.}~\bibnamefont
  {Thomas}},\ }\href {https://doi.org/https://doi.org/10.1007/BF01558908}
  {\bibfield  {journal} {\bibinfo  {journal} {Naturwissenschaften}\ }\textbf
  {\bibinfo {volume} {13}},\ \bibinfo {pages} {627} (\bibinfo {year}
  {1925})}\BibitemShut {NoStop}%
\bibitem [{\citenamefont {Kuhn}(1925)}]{Kuhn1925}%
  \BibitemOpen
  \bibfield  {author} {\bibinfo {author} {\bibfnamefont {W.}~\bibnamefont
  {Kuhn}},\ }\href {https://doi.org/https://doi.org/10.1007/BF01328322}
  {\bibfield  {journal} {\bibinfo  {journal} {Zeitschrift f{\"u}r Physik}\
  }\textbf {\bibinfo {volume} {33}},\ \bibinfo {pages} {408} (\bibinfo {year}
  {1925})}\BibitemShut {NoStop}%
\bibitem [{\citenamefont {Reiche}\ and\ \citenamefont
  {Thomas}(1925)}]{Reiche1925}%
  \BibitemOpen
  \bibfield  {author} {\bibinfo {author} {\bibfnamefont {F.}~\bibnamefont
  {Reiche}}\ and\ \bibinfo {author} {\bibfnamefont {W.}~\bibnamefont
  {Thomas}},\ }\href {https://doi.org/https://doi.org/10.1007/BF01328494}
  {\bibfield  {journal} {\bibinfo  {journal} {Zeitschrift f{\"u}r Physik}\
  }\textbf {\bibinfo {volume} {34}},\ \bibinfo {pages} {510} (\bibinfo {year}
  {1925})}\BibitemShut {NoStop}%
\bibitem [{\citenamefont {Bethe}(1930)}]{Bethe1930}%
  \BibitemOpen
  \bibfield  {author} {\bibinfo {author} {\bibfnamefont {H.}~\bibnamefont
  {Bethe}},\ }\href {https://doi.org/10.1002/andp.19303970303} {\bibfield
  {journal} {\bibinfo  {journal} {Ann. Phys.}\ }\textbf {\bibinfo {volume}
  {397}},\ \bibinfo {pages} {325} (\bibinfo {year} {1930})}\BibitemShut
  {NoStop}%
\bibitem [{\citenamefont {Wang}(1999)}]{Wang1999}%
  \BibitemOpen
  \bibfield  {author} {\bibinfo {author} {\bibfnamefont {S.}~\bibnamefont
  {Wang}},\ }\href {https://doi.org/https://doi.org/10.1103/PhysRevA.60.262}
  {\bibfield  {journal} {\bibinfo  {journal} {Phys. Rev. A}\ }\textbf {\bibinfo
  {volume} {60}},\ \bibinfo {pages} {262} (\bibinfo {year} {1999})}\BibitemShut
  {NoStop}%
\bibitem [{\citenamefont {Barnett}\ and\ \citenamefont
  {Loudon}(1996)}]{Barnett1996}%
  \BibitemOpen
  \bibfield  {author} {\bibinfo {author} {\bibfnamefont {S.~M.}\ \bibnamefont
  {Barnett}}\ and\ \bibinfo {author} {\bibfnamefont {R.}~\bibnamefont
  {Loudon}},\ }\href
  {https://doi.org/https://doi.org/10.1103/PhysRevLett.77.2444} {\bibfield
  {journal} {\bibinfo  {journal} {Phys. Rev. Lett.}\ }\textbf {\bibinfo
  {volume} {77}},\ \bibinfo {pages} {2444} (\bibinfo {year}
  {1996})}\BibitemShut {NoStop}%
\bibitem [{\citenamefont {Merzbacher}(1970)}]{Merzbacher1970}%
  \BibitemOpen
  \bibfield  {author} {\bibinfo {author} {\bibfnamefont {E.}~\bibnamefont
  {Merzbacher}},\ }\href@noop {} {\emph {\bibinfo {title} {Quantum
  Mechanics}}},\ \bibinfo {edition} {2nd}\ ed.\ (\bibinfo  {publisher} {Wiley,
  New York},\ \bibinfo {year} {1970})\BibitemShut {NoStop}%
\bibitem [{\citenamefont {Bassani}\ and\ \citenamefont
  {Scandolo}(1991)}]{Bassani1991}%
  \BibitemOpen
  \bibfield  {author} {\bibinfo {author} {\bibfnamefont {F.}~\bibnamefont
  {Bassani}}\ and\ \bibinfo {author} {\bibfnamefont {S.}~\bibnamefont
  {Scandolo}},\ }\href
  {https://doi.org/https://doi.org/10.1103/PhysRevB.44.8446} {\bibfield
  {journal} {\bibinfo  {journal} {Phys. Rev. B}\ }\textbf {\bibinfo {volume}
  {44}},\ \bibinfo {pages} {8446} (\bibinfo {year} {1991})}\BibitemShut
  {NoStop}%
\bibitem [{\citenamefont {Scandolo}\ and\ \citenamefont
  {Bassani}(1992)}]{Scandolo1992}%
  \BibitemOpen
  \bibfield  {author} {\bibinfo {author} {\bibfnamefont {S.}~\bibnamefont
  {Scandolo}}\ and\ \bibinfo {author} {\bibfnamefont {F.}~\bibnamefont
  {Bassani}},\ }\href
  {https://doi.org/https://doi.org/10.1103/PhysRevB.45.13257} {\bibfield
  {journal} {\bibinfo  {journal} {Phys. Rev. B}\ }\textbf {\bibinfo {volume}
  {45}},\ \bibinfo {pages} {13257} (\bibinfo {year} {1992})}\BibitemShut
  {NoStop}%
\bibitem [{\citenamefont {Scandolo}\ and\ \citenamefont
  {Bassani}(1995)}]{Scandolo1995}%
  \BibitemOpen
  \bibfield  {author} {\bibinfo {author} {\bibfnamefont {S.}~\bibnamefont
  {Scandolo}}\ and\ \bibinfo {author} {\bibfnamefont {F.}~\bibnamefont
  {Bassani}},\ }\href
  {https://doi.org/https://doi.org/10.1103/PhysRevB.51.6925} {\bibfield
  {journal} {\bibinfo  {journal} {Phys. Rev. B}\ }\textbf {\bibinfo {volume}
  {51}},\ \bibinfo {pages} {6925} (\bibinfo {year} {1995})}\BibitemShut
  {NoStop}%
\bibitem [{\citenamefont {Baxter}(1994)}]{Baxter1994}%
  \BibitemOpen
  \bibfield  {author} {\bibinfo {author} {\bibfnamefont {C.}~\bibnamefont
  {Baxter}},\ }\href {https://doi.org/10.1103/PhysRevA.50.875} {\bibfield
  {journal} {\bibinfo  {journal} {Phys. Rev. A}\ }\textbf {\bibinfo {volume}
  {50}},\ \bibinfo {pages} {875} (\bibinfo {year} {1994})}\BibitemShut
  {NoStop}%
\bibitem [{\citenamefont {Levinger}\ \emph {et~al.}(1957)\citenamefont
  {Levinger}, \citenamefont {Rustgi},\ and\ \citenamefont
  {Okamoto}}]{Levinger1957}%
  \BibitemOpen
  \bibfield  {author} {\bibinfo {author} {\bibfnamefont {J.~S.}\ \bibnamefont
  {Levinger}}, \bibinfo {author} {\bibfnamefont {M.~L.}\ \bibnamefont
  {Rustgi}},\ and\ \bibinfo {author} {\bibfnamefont {K.}~\bibnamefont
  {Okamoto}},\ }\href {https://doi.org/10.1103/PhysRev.106.1191} {\bibfield
  {journal} {\bibinfo  {journal} {Phys. Rev.}\ }\textbf {\bibinfo {volume}
  {106}},\ \bibinfo {pages} {1191} (\bibinfo {year} {1957})}\BibitemShut
  {NoStop}%
\bibitem [{\citenamefont {Friar}\ and\ \citenamefont
  {Fallieros}(1975)}]{Friar1975}%
  \BibitemOpen
  \bibfield  {author} {\bibinfo {author} {\bibfnamefont {J.~L.}\ \bibnamefont
  {Friar}}\ and\ \bibinfo {author} {\bibfnamefont {S.}~\bibnamefont
  {Fallieros}},\ }\href {https://doi.org/10.1103/PhysRevC.11.274} {\bibfield
  {journal} {\bibinfo  {journal} {Phys. Rev. C}\ }\textbf {\bibinfo {volume}
  {11}},\ \bibinfo {pages} {274} (\bibinfo {year} {1975})}\BibitemShut
  {NoStop}%
\bibitem [{\citenamefont {Nielsen}\ \emph {et~al.}(2010)\citenamefont
  {Nielsen}, \citenamefont {Navarra},\ and\ \citenamefont {Lee}}]{Nielsen2010}%
  \BibitemOpen
  \bibfield  {author} {\bibinfo {author} {\bibfnamefont {M.}~\bibnamefont
  {Nielsen}}, \bibinfo {author} {\bibfnamefont {F.~S.}\ \bibnamefont
  {Navarra}},\ and\ \bibinfo {author} {\bibfnamefont {S.~H.}\ \bibnamefont
  {Lee}},\ }\href {https://doi.org/10.1016/j.physrep.2010.07.005} {\bibfield
  {journal} {\bibinfo  {journal} {Phys. Rep.}\ }\textbf {\bibinfo {volume}
  {497}},\ \bibinfo {pages} {41} (\bibinfo {year} {2010})}\BibitemShut
  {NoStop}%
\bibitem [{\citenamefont {Pines}\ and\ \citenamefont
  {Nozieres}(1966)}]{Pines1966}%
  \BibitemOpen
  \bibfield  {author} {\bibinfo {author} {\bibfnamefont {D.}~\bibnamefont
  {Pines}}\ and\ \bibinfo {author} {\bibfnamefont {P.}~\bibnamefont
  {Nozieres}},\ }\href@noop {} {\emph {\bibinfo {title} {{The Theory of Quantum
  Liquids}}}}\ (\bibinfo  {publisher} {W. A. Benjamin},\ \bibinfo {address}
  {New York},\ \bibinfo {year} {1966})\BibitemShut {NoStop}%
\bibitem [{\citenamefont {Giuliani}\ and\ \citenamefont
  {Vignale}(2005)}]{Giuliani2005}%
  \BibitemOpen
  \bibfield  {author} {\bibinfo {author} {\bibfnamefont {G.}~\bibnamefont
  {Giuliani}}\ and\ \bibinfo {author} {\bibfnamefont {G.}~\bibnamefont
  {Vignale}},\ }\href@noop {} {\emph {\bibinfo {title} {Quantum theory of the
  electron liquid}}}\ (\bibinfo  {publisher} {Cambridge university press},\
  \bibinfo {year} {2005})\BibitemShut {NoStop}%
\bibitem [{\citenamefont {Andolina}\ \emph {et~al.}(2019)\citenamefont
  {Andolina}, \citenamefont {Pellegrino}, \citenamefont {Giovannetti},
  \citenamefont {MacDonald},\ and\ \citenamefont {Polini}}]{Andolina2019}%
  \BibitemOpen
  \bibfield  {author} {\bibinfo {author} {\bibfnamefont {G.~M.}\ \bibnamefont
  {Andolina}}, \bibinfo {author} {\bibfnamefont {F.~M.~D.}\ \bibnamefont
  {Pellegrino}}, \bibinfo {author} {\bibfnamefont {V.}~\bibnamefont
  {Giovannetti}}, \bibinfo {author} {\bibfnamefont {A.~H.}\ \bibnamefont
  {MacDonald}},\ and\ \bibinfo {author} {\bibfnamefont {M.}~\bibnamefont
  {Polini}},\ }\href {https://doi.org/10.1103/PhysRevB.100.121109} {\bibfield
  {journal} {\bibinfo  {journal} {Phys. Rev. B}\ }\textbf {\bibinfo {volume}
  {100}},\ \bibinfo {pages} {121109} (\bibinfo {year} {2019})}\BibitemShut
  {NoStop}%
\bibitem [{\citenamefont {Garziano}\ \emph {et~al.}(2020)\citenamefont
  {Garziano}, \citenamefont {Settineri}, \citenamefont {Di~Stefano},
  \citenamefont {Savasta},\ and\ \citenamefont {Nori}}]{Garziano2020}%
  \BibitemOpen
  \bibfield  {author} {\bibinfo {author} {\bibfnamefont {L.}~\bibnamefont
  {Garziano}}, \bibinfo {author} {\bibfnamefont {A.}~\bibnamefont {Settineri}},
  \bibinfo {author} {\bibfnamefont {O.}~\bibnamefont {Di~Stefano}}, \bibinfo
  {author} {\bibfnamefont {S.}~\bibnamefont {Savasta}},\ and\ \bibinfo {author}
  {\bibfnamefont {F.}~\bibnamefont {Nori}},\ }\href@noop {} {\bibfield
  {journal} {\bibinfo  {journal} {arXiv:2002.04241}\ } (\bibinfo {year}
  {2020})}\BibitemShut {NoStop}%
\bibitem [{\citenamefont {Anderson}(1958)}]{Anderson1958}%
  \BibitemOpen
  \bibfield  {author} {\bibinfo {author} {\bibfnamefont {P.~W.}\ \bibnamefont
  {Anderson}},\ }\href {https://doi.org/10.1103/PhysRev.112.1900} {\bibfield
  {journal} {\bibinfo  {journal} {Phys. Rev.}\ }\textbf {\bibinfo {volume}
  {112}},\ \bibinfo {pages} {1900} (\bibinfo {year} {1958})}\BibitemShut
  {NoStop}%
\bibitem [{\citenamefont {Barnett}\ and\ \citenamefont
  {Loudon}(2012)}]{Barnett2012}%
  \BibitemOpen
  \bibfield  {author} {\bibinfo {author} {\bibfnamefont {S.~M.}\ \bibnamefont
  {Barnett}}\ and\ \bibinfo {author} {\bibfnamefont {R.}~\bibnamefont
  {Loudon}},\ }\href
  {https://doi.org/https://doi.org/10.1103/PhysRevLett.108.013601} {\bibfield
  {journal} {\bibinfo  {journal} {Phys. Rev. Lett.}\ }\textbf {\bibinfo
  {volume} {108}},\ \bibinfo {pages} {013601} (\bibinfo {year}
  {2012})}\BibitemShut {NoStop}%
\bibitem [{\citenamefont {Kockum}\ \emph {et~al.}(2019)\citenamefont {Kockum},
  \citenamefont {Miranowicz}, \citenamefont {Liberato}, \citenamefont
  {Savasta},\ and\ \citenamefont {Nori}}]{Kockum2018}%
  \BibitemOpen
  \bibfield  {author} {\bibinfo {author} {\bibfnamefont {A.~F.}\ \bibnamefont
  {Kockum}}, \bibinfo {author} {\bibfnamefont {A.}~\bibnamefont {Miranowicz}},
  \bibinfo {author} {\bibfnamefont {S.~D.}\ \bibnamefont {Liberato}}, \bibinfo
  {author} {\bibfnamefont {S.}~\bibnamefont {Savasta}},\ and\ \bibinfo {author}
  {\bibfnamefont {F.}~\bibnamefont {Nori}},\ }\href
  {https://doi.org/10.1038/s42254-018-0006-2} {\bibfield  {journal} {\bibinfo
  {journal} {Nat. Rev. Phys.}\ }\textbf {\bibinfo {volume} {1}},\ \bibinfo
  {pages} {19} (\bibinfo {year} {2019})}\BibitemShut {NoStop}%
\bibitem [{\citenamefont {Forn-D{\'{i}}az}\ \emph {et~al.}(2019)\citenamefont
  {Forn-D{\'{i}}az}, \citenamefont {Lamata}, \citenamefont {Rico},
  \citenamefont {Kono},\ and\ \citenamefont {Solano}}]{Forn-Diaz2018}%
  \BibitemOpen
  \bibfield  {author} {\bibinfo {author} {\bibfnamefont {P.}~\bibnamefont
  {Forn-D{\'{i}}az}}, \bibinfo {author} {\bibfnamefont {L.}~\bibnamefont
  {Lamata}}, \bibinfo {author} {\bibfnamefont {E.}~\bibnamefont {Rico}},
  \bibinfo {author} {\bibfnamefont {J.}~\bibnamefont {Kono}},\ and\ \bibinfo
  {author} {\bibfnamefont {E.}~\bibnamefont {Solano}},\ }\href
  {https://doi.org/10.1103/RevModPhys.91.025005} {\bibfield  {journal}
  {\bibinfo  {journal} {Rev. Mod. Phys.}\ }\textbf {\bibinfo {volume} {91}},\
  \bibinfo {pages} {025005} (\bibinfo {year} {2019})}\BibitemShut {NoStop}%
\bibitem [{\citenamefont {Peyronel}\ \emph {et~al.}(2012)\citenamefont
  {Peyronel}, \citenamefont {Firstenberg}, \citenamefont {Liang}, \citenamefont
  {Hofferberth}, \citenamefont {Gorshkov}, \citenamefont {Pohl}, \citenamefont
  {Lukin},\ and\ \citenamefont {Vuleti{\'c}}}]{Peyronel2012}%
  \BibitemOpen
  \bibfield  {author} {\bibinfo {author} {\bibfnamefont {T.}~\bibnamefont
  {Peyronel}}, \bibinfo {author} {\bibfnamefont {O.}~\bibnamefont
  {Firstenberg}}, \bibinfo {author} {\bibfnamefont {Q.-Y.}\ \bibnamefont
  {Liang}}, \bibinfo {author} {\bibfnamefont {S.}~\bibnamefont {Hofferberth}},
  \bibinfo {author} {\bibfnamefont {A.~V.}\ \bibnamefont {Gorshkov}}, \bibinfo
  {author} {\bibfnamefont {T.}~\bibnamefont {Pohl}}, \bibinfo {author}
  {\bibfnamefont {M.~D.}\ \bibnamefont {Lukin}},\ and\ \bibinfo {author}
  {\bibfnamefont {V.}~\bibnamefont {Vuleti{\'c}}},\ }\href
  {https://doi.org/https://doi.org/10.1038/nature11361} {\bibfield  {journal}
  {\bibinfo  {journal} {Nature}\ }\textbf {\bibinfo {volume} {488}},\ \bibinfo
  {pages} {57} (\bibinfo {year} {2012})}\BibitemShut {NoStop}%
\bibitem [{\citenamefont {Chang}\ \emph {et~al.}(2014)\citenamefont {Chang},
  \citenamefont {Vuleti{\'c}},\ and\ \citenamefont {Lukin}}]{Chang2014}%
  \BibitemOpen
  \bibfield  {author} {\bibinfo {author} {\bibfnamefont {D.~E.}\ \bibnamefont
  {Chang}}, \bibinfo {author} {\bibfnamefont {V.}~\bibnamefont {Vuleti{\'c}}},\
  and\ \bibinfo {author} {\bibfnamefont {M.~D.}\ \bibnamefont {Lukin}},\ }\href
  {https://doi.org/10.1038/nphoton.2014.192} {\bibfield  {journal} {\bibinfo
  {journal} {Nat. Photonics}\ }\textbf {\bibinfo {volume} {8}},\ \bibinfo
  {pages} {685} (\bibinfo {year} {2014})}\BibitemShut {NoStop}%
\bibitem [{\citenamefont {Guerreiro}\ \emph {et~al.}(2014)\citenamefont
  {Guerreiro}, \citenamefont {Martin}, \citenamefont {Sanguinetti},
  \citenamefont {Pelc}, \citenamefont {Langrock}, \citenamefont {Fejer},
  \citenamefont {Gisin}, \citenamefont {Zbinden}, \citenamefont {Sangouard},\
  and\ \citenamefont {Thew}}]{Guerreiro2014}%
  \BibitemOpen
  \bibfield  {author} {\bibinfo {author} {\bibfnamefont {T.}~\bibnamefont
  {Guerreiro}}, \bibinfo {author} {\bibfnamefont {A.}~\bibnamefont {Martin}},
  \bibinfo {author} {\bibfnamefont {B.}~\bibnamefont {Sanguinetti}}, \bibinfo
  {author} {\bibfnamefont {J.}~\bibnamefont {Pelc}}, \bibinfo {author}
  {\bibfnamefont {C.}~\bibnamefont {Langrock}}, \bibinfo {author}
  {\bibfnamefont {M.}~\bibnamefont {Fejer}}, \bibinfo {author} {\bibfnamefont
  {N.}~\bibnamefont {Gisin}}, \bibinfo {author} {\bibfnamefont
  {H.}~\bibnamefont {Zbinden}}, \bibinfo {author} {\bibfnamefont
  {N.}~\bibnamefont {Sangouard}},\ and\ \bibinfo {author} {\bibfnamefont
  {R.}~\bibnamefont {Thew}},\ }\href
  {https://doi.org/https://doi.org/10.1103/PhysRevLett.113.173601} {\bibfield
  {journal} {\bibinfo  {journal} {Phys. Rev. Lett.}\ }\textbf {\bibinfo
  {volume} {113}},\ \bibinfo {pages} {173601} (\bibinfo {year}
  {2014})}\BibitemShut {NoStop}%
\bibitem [{\citenamefont {Kockum}\ \emph
  {et~al.}(2017{\natexlab{a}})\citenamefont {Kockum}, \citenamefont
  {Miranowicz}, \citenamefont {Macr{\`{i}}}, \citenamefont {Savasta},\ and\
  \citenamefont {Nori}}]{Kockum2017a}%
  \BibitemOpen
  \bibfield  {author} {\bibinfo {author} {\bibfnamefont {A.~F.}\ \bibnamefont
  {Kockum}}, \bibinfo {author} {\bibfnamefont {A.}~\bibnamefont {Miranowicz}},
  \bibinfo {author} {\bibfnamefont {V.}~\bibnamefont {Macr{\`{i}}}}, \bibinfo
  {author} {\bibfnamefont {S.}~\bibnamefont {Savasta}},\ and\ \bibinfo {author}
  {\bibfnamefont {F.}~\bibnamefont {Nori}},\ }\href
  {https://doi.org/10.1103/PhysRevA.95.063849} {\bibfield  {journal} {\bibinfo
  {journal} {Phys. Rev. A}\ }\textbf {\bibinfo {volume} {95}},\ \bibinfo
  {pages} {063849} (\bibinfo {year} {2017}{\natexlab{a}})}\BibitemShut
  {NoStop}%
\bibitem [{\citenamefont {Carusotto}\ and\ \citenamefont
  {Ciuti}(2013)}]{Carusotto2013}%
  \BibitemOpen
  \bibfield  {author} {\bibinfo {author} {\bibfnamefont {I.}~\bibnamefont
  {Carusotto}}\ and\ \bibinfo {author} {\bibfnamefont {C.}~\bibnamefont
  {Ciuti}},\ }\href {https://doi.org/10.1103/RevModPhys.85.299} {\bibfield
  {journal} {\bibinfo  {journal} {Rev. Mod. Phys.}\ }\textbf {\bibinfo {volume}
  {85}},\ \bibinfo {pages} {299} (\bibinfo {year} {2013})}\BibitemShut
  {NoStop}%
\bibitem [{\citenamefont {{De Liberato}}(2014)}]{DeLiberato2014}%
  \BibitemOpen
  \bibfield  {author} {\bibinfo {author} {\bibfnamefont {S.}~\bibnamefont {{De
  Liberato}}},\ }\href {https://doi.org/10.1103/PhysRevLett.112.016401}
  {\bibfield  {journal} {\bibinfo  {journal} {Phys. Rev. Lett.}\ }\textbf
  {\bibinfo {volume} {112}},\ \bibinfo {pages} {016401} (\bibinfo {year}
  {2014})}\BibitemShut {NoStop}%
\bibitem [{\citenamefont {Garc{\'{i}}a-Ripoll}\ \emph
  {et~al.}(2015)\citenamefont {Garc{\'{i}}a-Ripoll}, \citenamefont
  {Peropadre},\ and\ \citenamefont {{De Liberato}}}]{Garcia-Ripoll2015}%
  \BibitemOpen
  \bibfield  {author} {\bibinfo {author} {\bibfnamefont {J.~J.}\ \bibnamefont
  {Garc{\'{i}}a-Ripoll}}, \bibinfo {author} {\bibfnamefont {B.}~\bibnamefont
  {Peropadre}},\ and\ \bibinfo {author} {\bibfnamefont {S.}~\bibnamefont {{De
  Liberato}}},\ }\href {https://doi.org/10.1038/srep16055} {\bibfield
  {journal} {\bibinfo  {journal} {Sci. Rep.}\ }\textbf {\bibinfo {volume}
  {5}},\ \bibinfo {pages} {16055} (\bibinfo {year} {2015})}\BibitemShut
  {NoStop}%
\bibitem [{\citenamefont {Bayer}\ \emph {et~al.}(2017)\citenamefont {Bayer},
  \citenamefont {Pozimski}, \citenamefont {Schambeck}, \citenamefont {Schuh},
  \citenamefont {Huber}, \citenamefont {Bougeard},\ and\ \citenamefont
  {Lange}}]{Bayer2017}%
  \BibitemOpen
  \bibfield  {author} {\bibinfo {author} {\bibfnamefont {A.}~\bibnamefont
  {Bayer}}, \bibinfo {author} {\bibfnamefont {M.}~\bibnamefont {Pozimski}},
  \bibinfo {author} {\bibfnamefont {S.}~\bibnamefont {Schambeck}}, \bibinfo
  {author} {\bibfnamefont {D.}~\bibnamefont {Schuh}}, \bibinfo {author}
  {\bibfnamefont {R.}~\bibnamefont {Huber}}, \bibinfo {author} {\bibfnamefont
  {D.}~\bibnamefont {Bougeard}},\ and\ \bibinfo {author} {\bibfnamefont
  {C.}~\bibnamefont {Lange}},\ }\href
  {https://doi.org/10.1021/acs.nanolett.7b03103} {\bibfield  {journal}
  {\bibinfo  {journal} {Nano Lett.}\ }\textbf {\bibinfo {volume} {17}},\
  \bibinfo {pages} {6340} (\bibinfo {year} {2017})}\BibitemShut {NoStop}%
\bibitem [{\citenamefont {Yoshihara}\ \emph {et~al.}(2017)\citenamefont
  {Yoshihara}, \citenamefont {Fuse}, \citenamefont {Ashhab}, \citenamefont
  {Kakuyanagi}, \citenamefont {Saito},\ and\ \citenamefont
  {Semba}}]{Yoshihara2017a}%
  \BibitemOpen
  \bibfield  {author} {\bibinfo {author} {\bibfnamefont {F.}~\bibnamefont
  {Yoshihara}}, \bibinfo {author} {\bibfnamefont {T.}~\bibnamefont {Fuse}},
  \bibinfo {author} {\bibfnamefont {S.}~\bibnamefont {Ashhab}}, \bibinfo
  {author} {\bibfnamefont {K.}~\bibnamefont {Kakuyanagi}}, \bibinfo {author}
  {\bibfnamefont {S.}~\bibnamefont {Saito}},\ and\ \bibinfo {author}
  {\bibfnamefont {K.}~\bibnamefont {Semba}},\ }\href
  {https://doi.org/10.1103/PhysRevA.95.053824} {\bibfield  {journal} {\bibinfo
  {journal} {Phys. Rev. A}\ }\textbf {\bibinfo {volume} {95}},\ \bibinfo
  {pages} {053824} (\bibinfo {year} {2017})}\BibitemShut {NoStop}%
\bibitem [{\citenamefont {Ridolfo}\ \emph {et~al.}(2012)\citenamefont
  {Ridolfo}, \citenamefont {Leib}, \citenamefont {Savasta},\ and\ \citenamefont
  {Hartmann}}]{Ridolfo2012}%
  \BibitemOpen
  \bibfield  {author} {\bibinfo {author} {\bibfnamefont {A.}~\bibnamefont
  {Ridolfo}}, \bibinfo {author} {\bibfnamefont {M.}~\bibnamefont {Leib}},
  \bibinfo {author} {\bibfnamefont {S.}~\bibnamefont {Savasta}},\ and\ \bibinfo
  {author} {\bibfnamefont {M.~J.}\ \bibnamefont {Hartmann}},\ }\href
  {https://doi.org/10.1103/PhysRevLett.109.193602} {\bibfield  {journal}
  {\bibinfo  {journal} {Phys. Rev. Lett.}\ }\textbf {\bibinfo {volume} {109}},\
  \bibinfo {pages} {193602} (\bibinfo {year} {2012})}\BibitemShut {NoStop}%
\bibitem [{\citenamefont {Niemczyk}\ \emph {et~al.}(2010)\citenamefont
  {Niemczyk}, \citenamefont {Deppe}, \citenamefont {Huebl}, \citenamefont
  {Menzel}, \citenamefont {Hocke}, \citenamefont {Schwarz}, \citenamefont
  {Garcia-Ripoll}, \citenamefont {Zueco}, \citenamefont {H{\"{u}}mmer},
  \citenamefont {Solano}, \citenamefont {Marx},\ and\ \citenamefont
  {Gross}}]{Niemczyk2010}%
  \BibitemOpen
  \bibfield  {author} {\bibinfo {author} {\bibfnamefont {T.}~\bibnamefont
  {Niemczyk}}, \bibinfo {author} {\bibfnamefont {F.}~\bibnamefont {Deppe}},
  \bibinfo {author} {\bibfnamefont {H.}~\bibnamefont {Huebl}}, \bibinfo
  {author} {\bibfnamefont {E.~P.}\ \bibnamefont {Menzel}}, \bibinfo {author}
  {\bibfnamefont {F.}~\bibnamefont {Hocke}}, \bibinfo {author} {\bibfnamefont
  {M.~J.}\ \bibnamefont {Schwarz}}, \bibinfo {author} {\bibfnamefont {J.~J.}\
  \bibnamefont {Garcia-Ripoll}}, \bibinfo {author} {\bibfnamefont
  {D.}~\bibnamefont {Zueco}}, \bibinfo {author} {\bibfnamefont
  {T.}~\bibnamefont {H{\"{u}}mmer}}, \bibinfo {author} {\bibfnamefont
  {E.}~\bibnamefont {Solano}}, \bibinfo {author} {\bibfnamefont
  {A.}~\bibnamefont {Marx}},\ and\ \bibinfo {author} {\bibfnamefont
  {R.}~\bibnamefont {Gross}},\ }\href {https://doi.org/10.1038/nphys1730}
  {\bibfield  {journal} {\bibinfo  {journal} {Nat. Phys.}\ }\textbf {\bibinfo
  {volume} {6}},\ \bibinfo {pages} {772} (\bibinfo {year} {2010})}\BibitemShut
  {NoStop}%
\bibitem [{\citenamefont {Casanova}\ \emph {et~al.}(2010)\citenamefont
  {Casanova}, \citenamefont {Romero}, \citenamefont {Lizuain}, \citenamefont
  {Garc{\'{i}}a-Ripoll},\ and\ \citenamefont {Solano}}]{Casanova2010}%
  \BibitemOpen
  \bibfield  {author} {\bibinfo {author} {\bibfnamefont {J.}~\bibnamefont
  {Casanova}}, \bibinfo {author} {\bibfnamefont {G.}~\bibnamefont {Romero}},
  \bibinfo {author} {\bibfnamefont {I.}~\bibnamefont {Lizuain}}, \bibinfo
  {author} {\bibfnamefont {J.~J.}\ \bibnamefont {Garc{\'{i}}a-Ripoll}},\ and\
  \bibinfo {author} {\bibfnamefont {E.}~\bibnamefont {Solano}},\ }\href
  {https://doi.org/10.1103/PhysRevLett.105.263603} {\bibfield  {journal}
  {\bibinfo  {journal} {Phys. Rev. Lett.}\ }\textbf {\bibinfo {volume} {105}},\
  \bibinfo {pages} {263603} (\bibinfo {year} {2010})}\BibitemShut {NoStop}%
\bibitem [{\citenamefont {Garziano}\ \emph {et~al.}(2016)\citenamefont
  {Garziano}, \citenamefont {Macr{\`{i}}}, \citenamefont {Stassi},
  \citenamefont {{Di Stefano}}, \citenamefont {Nori},\ and\ \citenamefont
  {Savasta}}]{Garziano2016}%
  \BibitemOpen
  \bibfield  {author} {\bibinfo {author} {\bibfnamefont {L.}~\bibnamefont
  {Garziano}}, \bibinfo {author} {\bibfnamefont {V.}~\bibnamefont
  {Macr{\`{i}}}}, \bibinfo {author} {\bibfnamefont {R.}~\bibnamefont {Stassi}},
  \bibinfo {author} {\bibfnamefont {O.}~\bibnamefont {{Di Stefano}}}, \bibinfo
  {author} {\bibfnamefont {F.}~\bibnamefont {Nori}},\ and\ \bibinfo {author}
  {\bibfnamefont {S.}~\bibnamefont {Savasta}},\ }\href
  {https://doi.org/10.1103/PhysRevLett.117.043601} {\bibfield  {journal}
  {\bibinfo  {journal} {Phys. Rev. Lett.}\ }\textbf {\bibinfo {volume} {117}},\
  \bibinfo {pages} {043601} (\bibinfo {year} {2016})}\BibitemShut {NoStop}%
\bibitem [{\citenamefont {Stassi}\ \emph {et~al.}(2017)\citenamefont {Stassi},
  \citenamefont {Macr{\`{i}}}, \citenamefont {Kockum}, \citenamefont {{Di
  Stefano}}, \citenamefont {Miranowicz}, \citenamefont {Savasta},\ and\
  \citenamefont {Nori}}]{Stassi2017}%
  \BibitemOpen
  \bibfield  {author} {\bibinfo {author} {\bibfnamefont {R.}~\bibnamefont
  {Stassi}}, \bibinfo {author} {\bibfnamefont {V.}~\bibnamefont {Macr{\`{i}}}},
  \bibinfo {author} {\bibfnamefont {A.~F.}\ \bibnamefont {Kockum}}, \bibinfo
  {author} {\bibfnamefont {O.}~\bibnamefont {{Di Stefano}}}, \bibinfo {author}
  {\bibfnamefont {A.}~\bibnamefont {Miranowicz}}, \bibinfo {author}
  {\bibfnamefont {S.}~\bibnamefont {Savasta}},\ and\ \bibinfo {author}
  {\bibfnamefont {F.}~\bibnamefont {Nori}},\ }\href
  {https://doi.org/10.1103/PhysRevA.96.023818} {\bibfield  {journal} {\bibinfo
  {journal} {Phys. Rev. A}\ }\textbf {\bibinfo {volume} {96}},\ \bibinfo
  {pages} {023818} (\bibinfo {year} {2017})}\BibitemShut {NoStop}%
\bibitem [{\citenamefont {{De Liberato}}\ \emph {et~al.}(2007)\citenamefont
  {{De Liberato}}, \citenamefont {Ciuti},\ and\ \citenamefont
  {Carusotto}}]{DeLiberato2007}%
  \BibitemOpen
  \bibfield  {author} {\bibinfo {author} {\bibfnamefont {S.}~\bibnamefont {{De
  Liberato}}}, \bibinfo {author} {\bibfnamefont {C.}~\bibnamefont {Ciuti}},\
  and\ \bibinfo {author} {\bibfnamefont {I.}~\bibnamefont {Carusotto}},\ }\href
  {https://doi.org/10.1103/PhysRevLett.98.103602} {\bibfield  {journal}
  {\bibinfo  {journal} {Phys. Rev. Lett.}\ }\textbf {\bibinfo {volume} {98}},\
  \bibinfo {pages} {103602} (\bibinfo {year} {2007})}\BibitemShut {NoStop}%
\bibitem [{\citenamefont {Ashhab}\ and\ \citenamefont
  {Nori}(2010)}]{Ashhab2010}%
  \BibitemOpen
  \bibfield  {author} {\bibinfo {author} {\bibfnamefont {S.}~\bibnamefont
  {Ashhab}}\ and\ \bibinfo {author} {\bibfnamefont {F.}~\bibnamefont {Nori}},\
  }\href {https://doi.org/10.1103/PhysRevA.81.042311} {\bibfield  {journal}
  {\bibinfo  {journal} {Phys. Rev. A}\ }\textbf {\bibinfo {volume} {81}},\
  \bibinfo {pages} {042311} (\bibinfo {year} {2010})}\BibitemShut {NoStop}%
\bibitem [{\citenamefont {Carusotto}\ \emph {et~al.}(2012)\citenamefont
  {Carusotto}, \citenamefont {{De Liberato}}, \citenamefont {Gerace},\ and\
  \citenamefont {Ciuti}}]{Carusotto2012}%
  \BibitemOpen
  \bibfield  {author} {\bibinfo {author} {\bibfnamefont {I.}~\bibnamefont
  {Carusotto}}, \bibinfo {author} {\bibfnamefont {S.}~\bibnamefont {{De
  Liberato}}}, \bibinfo {author} {\bibfnamefont {D.}~\bibnamefont {Gerace}},\
  and\ \bibinfo {author} {\bibfnamefont {C.}~\bibnamefont {Ciuti}},\ }\href
  {https://doi.org/10.1103/PhysRevA.85.023805} {\bibfield  {journal} {\bibinfo
  {journal} {Phys. Rev. A}\ }\textbf {\bibinfo {volume} {85}},\ \bibinfo
  {pages} {023805} (\bibinfo {year} {2012})}\BibitemShut {NoStop}%
\bibitem [{\citenamefont {Auer}\ and\ \citenamefont
  {Burkard}(2012)}]{Auer2012}%
  \BibitemOpen
  \bibfield  {author} {\bibinfo {author} {\bibfnamefont {A.}~\bibnamefont
  {Auer}}\ and\ \bibinfo {author} {\bibfnamefont {G.}~\bibnamefont {Burkard}},\
  }\href {https://doi.org/10.1103/PhysRevB.85.235140} {\bibfield  {journal}
  {\bibinfo  {journal} {Phys. Rev. B}\ }\textbf {\bibinfo {volume} {85}},\
  \bibinfo {pages} {235140} (\bibinfo {year} {2012})}\BibitemShut {NoStop}%
\bibitem [{\citenamefont {Garziano}\ \emph {et~al.}(2013)\citenamefont
  {Garziano}, \citenamefont {Ridolfo}, \citenamefont {Stassi}, \citenamefont
  {{Di Stefano}},\ and\ \citenamefont {Savasta}}]{Garziano2013}%
  \BibitemOpen
  \bibfield  {author} {\bibinfo {author} {\bibfnamefont {L.}~\bibnamefont
  {Garziano}}, \bibinfo {author} {\bibfnamefont {A.}~\bibnamefont {Ridolfo}},
  \bibinfo {author} {\bibfnamefont {R.}~\bibnamefont {Stassi}}, \bibinfo
  {author} {\bibfnamefont {O.}~\bibnamefont {{Di Stefano}}},\ and\ \bibinfo
  {author} {\bibfnamefont {S.}~\bibnamefont {Savasta}},\ }\href
  {https://doi.org/10.1103/PhysRevA.88.063829} {\bibfield  {journal} {\bibinfo
  {journal} {Phys. Rev. A}\ }\textbf {\bibinfo {volume} {88}},\ \bibinfo
  {pages} {063829} (\bibinfo {year} {2013})}\BibitemShut {NoStop}%
\bibitem [{\citenamefont {Stassi}\ \emph {et~al.}(2013)\citenamefont {Stassi},
  \citenamefont {Ridolfo}, \citenamefont {{Di Stefano}}, \citenamefont
  {Hartmann},\ and\ \citenamefont {Savasta}}]{Stassi2013}%
  \BibitemOpen
  \bibfield  {author} {\bibinfo {author} {\bibfnamefont {R.}~\bibnamefont
  {Stassi}}, \bibinfo {author} {\bibfnamefont {A.}~\bibnamefont {Ridolfo}},
  \bibinfo {author} {\bibfnamefont {O.}~\bibnamefont {{Di Stefano}}}, \bibinfo
  {author} {\bibfnamefont {M.~J.}\ \bibnamefont {Hartmann}},\ and\ \bibinfo
  {author} {\bibfnamefont {S.}~\bibnamefont {Savasta}},\ }\href
  {https://doi.org/10.1103/PhysRevLett.110.243601} {\bibfield  {journal}
  {\bibinfo  {journal} {Phys. Rev. Lett.}\ }\textbf {\bibinfo {volume} {110}},\
  \bibinfo {pages} {243601} (\bibinfo {year} {2013})}\BibitemShut {NoStop}%
\bibitem [{\citenamefont {Garziano}\ \emph {et~al.}(2014)\citenamefont
  {Garziano}, \citenamefont {Stassi}, \citenamefont {Ridolfo}, \citenamefont
  {{Di Stefano}},\ and\ \citenamefont {Savasta}}]{Garziano2014}%
  \BibitemOpen
  \bibfield  {author} {\bibinfo {author} {\bibfnamefont {L.}~\bibnamefont
  {Garziano}}, \bibinfo {author} {\bibfnamefont {R.}~\bibnamefont {Stassi}},
  \bibinfo {author} {\bibfnamefont {A.}~\bibnamefont {Ridolfo}}, \bibinfo
  {author} {\bibfnamefont {O.}~\bibnamefont {{Di Stefano}}},\ and\ \bibinfo
  {author} {\bibfnamefont {S.}~\bibnamefont {Savasta}},\ }\href
  {https://doi.org/10.1103/PhysRevA.90.043817} {\bibfield  {journal} {\bibinfo
  {journal} {Phys. Rev. A}\ }\textbf {\bibinfo {volume} {90}},\ \bibinfo
  {pages} {043817} (\bibinfo {year} {2014})}\BibitemShut {NoStop}%
\bibitem [{\citenamefont {Huang}\ and\ \citenamefont {Law}(2014)}]{Huang2014}%
  \BibitemOpen
  \bibfield  {author} {\bibinfo {author} {\bibfnamefont {J.-F.}\ \bibnamefont
  {Huang}}\ and\ \bibinfo {author} {\bibfnamefont {C.~K.}\ \bibnamefont
  {Law}},\ }\href {https://doi.org/10.1103/PhysRevA.89.033827} {\bibfield
  {journal} {\bibinfo  {journal} {Phys. Rev. A}\ }\textbf {\bibinfo {volume}
  {89}},\ \bibinfo {pages} {033827} (\bibinfo {year} {2014})}\BibitemShut
  {NoStop}%
\bibitem [{\citenamefont {Benenti}\ \emph {et~al.}(2014)\citenamefont
  {Benenti}, \citenamefont {D'Arrigo}, \citenamefont {Siccardi},\ and\
  \citenamefont {Strini}}]{Benenti2014}%
  \BibitemOpen
  \bibfield  {author} {\bibinfo {author} {\bibfnamefont {G.}~\bibnamefont
  {Benenti}}, \bibinfo {author} {\bibfnamefont {A.}~\bibnamefont {D'Arrigo}},
  \bibinfo {author} {\bibfnamefont {S.}~\bibnamefont {Siccardi}},\ and\
  \bibinfo {author} {\bibfnamefont {G.}~\bibnamefont {Strini}},\ }\href
  {https://doi.org/https://doi.org/10.1103/PhysRevA.90.052313} {\bibfield
  {journal} {\bibinfo  {journal} {Phys. Rev. A}\ }\textbf {\bibinfo {volume}
  {90}},\ \bibinfo {pages} {052313} (\bibinfo {year} {2014})}\BibitemShut
  {NoStop}%
\bibitem [{\citenamefont {Garziano}\ \emph {et~al.}(2015)\citenamefont
  {Garziano}, \citenamefont {Stassi}, \citenamefont {Macr{\`{i}}},
  \citenamefont {Kockum}, \citenamefont {Savasta},\ and\ \citenamefont
  {Nori}}]{Garziano2015}%
  \BibitemOpen
  \bibfield  {author} {\bibinfo {author} {\bibfnamefont {L.}~\bibnamefont
  {Garziano}}, \bibinfo {author} {\bibfnamefont {R.}~\bibnamefont {Stassi}},
  \bibinfo {author} {\bibfnamefont {V.}~\bibnamefont {Macr{\`{i}}}}, \bibinfo
  {author} {\bibfnamefont {A.~F.}\ \bibnamefont {Kockum}}, \bibinfo {author}
  {\bibfnamefont {S.}~\bibnamefont {Savasta}},\ and\ \bibinfo {author}
  {\bibfnamefont {F.}~\bibnamefont {Nori}},\ }\href
  {https://doi.org/10.1103/PhysRevA.92.063830} {\bibfield  {journal} {\bibinfo
  {journal} {Phys. Rev. A}\ }\textbf {\bibinfo {volume} {92}},\ \bibinfo
  {pages} {063830} (\bibinfo {year} {2015})}\BibitemShut {NoStop}%
\bibitem [{\citenamefont {Stassi}\ \emph {et~al.}(2016)\citenamefont {Stassi},
  \citenamefont {Savasta}, \citenamefont {Garziano}, \citenamefont {Spagnolo},\
  and\ \citenamefont {Nori}}]{Stassi2016}%
  \BibitemOpen
  \bibfield  {author} {\bibinfo {author} {\bibfnamefont {R.}~\bibnamefont
  {Stassi}}, \bibinfo {author} {\bibfnamefont {S.}~\bibnamefont {Savasta}},
  \bibinfo {author} {\bibfnamefont {L.}~\bibnamefont {Garziano}}, \bibinfo
  {author} {\bibfnamefont {B.}~\bibnamefont {Spagnolo}},\ and\ \bibinfo
  {author} {\bibfnamefont {F.}~\bibnamefont {Nori}},\ }\href
  {https://doi.org/10.1088/1367-2630/18/12/123005} {\bibfield  {journal}
  {\bibinfo  {journal} {New J. Phys.}\ }\textbf {\bibinfo {volume} {18}},\
  \bibinfo {pages} {123005} (\bibinfo {year} {2016})}\BibitemShut {NoStop}%
\bibitem [{\citenamefont {Jaako}\ \emph {et~al.}(2016)\citenamefont {Jaako},
  \citenamefont {Xiang}, \citenamefont {Garcia-Ripoll},\ and\ \citenamefont
  {Rabl}}]{Jaako2016}%
  \BibitemOpen
  \bibfield  {author} {\bibinfo {author} {\bibfnamefont {T.}~\bibnamefont
  {Jaako}}, \bibinfo {author} {\bibfnamefont {Z.-L.}\ \bibnamefont {Xiang}},
  \bibinfo {author} {\bibfnamefont {J.~J.}\ \bibnamefont {Garcia-Ripoll}},\
  and\ \bibinfo {author} {\bibfnamefont {P.}~\bibnamefont {Rabl}},\ }\href
  {https://doi.org/10.1103/PhysRevA.94.033850} {\bibfield  {journal} {\bibinfo
  {journal} {Phys. Rev. A}\ }\textbf {\bibinfo {volume} {94}},\ \bibinfo
  {pages} {033850} (\bibinfo {year} {2016})}\BibitemShut {NoStop}%
\bibitem [{\citenamefont {De~Liberato}(2017)}]{DeLiberato2017}%
  \BibitemOpen
  \bibfield  {author} {\bibinfo {author} {\bibfnamefont {S.}~\bibnamefont
  {De~Liberato}},\ }\href {https://doi.org/10.1038/s41467-017-01504-5}
  {\bibfield  {journal} {\bibinfo  {journal} {Nat. Commun.}\ }\textbf {\bibinfo
  {volume} {8}},\ \bibinfo {pages} {1465} (\bibinfo {year} {2017})}\BibitemShut
  {NoStop}%
\bibitem [{\citenamefont {Cirio}\ \emph {et~al.}(2017)\citenamefont {Cirio},
  \citenamefont {Debnath}, \citenamefont {Lambert},\ and\ \citenamefont
  {Nori}}]{Cirio2017}%
  \BibitemOpen
  \bibfield  {author} {\bibinfo {author} {\bibfnamefont {M.}~\bibnamefont
  {Cirio}}, \bibinfo {author} {\bibfnamefont {K.}~\bibnamefont {Debnath}},
  \bibinfo {author} {\bibfnamefont {N.}~\bibnamefont {Lambert}},\ and\ \bibinfo
  {author} {\bibfnamefont {F.}~\bibnamefont {Nori}},\ }\href
  {https://doi.org/10.1103/PhysRevLett.119.053601} {\bibfield  {journal}
  {\bibinfo  {journal} {Phys. Rev. Lett.}\ }\textbf {\bibinfo {volume} {119}},\
  \bibinfo {pages} {053601} (\bibinfo {year} {2017})}\BibitemShut {NoStop}%
\bibitem [{\citenamefont {Kockum}\ \emph
  {et~al.}(2017{\natexlab{b}})\citenamefont {Kockum}, \citenamefont
  {Macr{\`{i}}}, \citenamefont {Garziano}, \citenamefont {Savasta},\ and\
  \citenamefont {Nori}}]{Kockum2017}%
  \BibitemOpen
  \bibfield  {author} {\bibinfo {author} {\bibfnamefont {A.~F.}\ \bibnamefont
  {Kockum}}, \bibinfo {author} {\bibfnamefont {V.}~\bibnamefont {Macr{\`{i}}}},
  \bibinfo {author} {\bibfnamefont {L.}~\bibnamefont {Garziano}}, \bibinfo
  {author} {\bibfnamefont {S.}~\bibnamefont {Savasta}},\ and\ \bibinfo {author}
  {\bibfnamefont {F.}~\bibnamefont {Nori}},\ }\href
  {https://doi.org/10.1038/s41598-017-04225-3} {\bibfield  {journal} {\bibinfo
  {journal} {Sci. Rep.}\ }\textbf {\bibinfo {volume} {7}},\ \bibinfo {pages}
  {5313} (\bibinfo {year} {2017}{\natexlab{b}})}\BibitemShut {NoStop}%
\bibitem [{\citenamefont {Albarr{\'{a}}n-Arriagada}\ \emph
  {et~al.}(2017)\citenamefont {Albarr{\'{a}}n-Arriagada}, \citenamefont {G.},
  \citenamefont {C{\'{a}}rdenas-L{\'{o}}pez}, \citenamefont {Romero},\ and\
  \citenamefont {Retamal}}]{Albarr_n_Arriagada_2017}%
  \BibitemOpen
  \bibfield  {author} {\bibinfo {author} {\bibfnamefont {F.}~\bibnamefont
  {Albarr{\'{a}}n-Arriagada}}, \bibinfo {author} {\bibfnamefont {A.~B.}\
  \bibnamefont {G.}}, \bibinfo {author} {\bibfnamefont {F.}~\bibnamefont
  {C{\'{a}}rdenas-L{\'{o}}pez}}, \bibinfo {author} {\bibfnamefont
  {G.}~\bibnamefont {Romero}},\ and\ \bibinfo {author} {\bibfnamefont {J.~C.}\
  \bibnamefont {Retamal}},\ }\href {https://doi.org/10.1088/1751-8121/aa66a0}
  {\bibfield  {journal} {\bibinfo  {journal} {J. Phys. A: Math. Theor.}\
  }\textbf {\bibinfo {volume} {50}},\ \bibinfo {pages} {184001} (\bibinfo
  {year} {2017})}\BibitemShut {NoStop}%
\bibitem [{\citenamefont {{Di Stefano}}\ \emph {et~al.}(2017)\citenamefont {{Di
  Stefano}}, \citenamefont {Stassi}, \citenamefont {Garziano}, \citenamefont
  {Kockum}, \citenamefont {Savasta},\ and\ \citenamefont
  {Nori}}]{DiStefano2017}%
  \BibitemOpen
  \bibfield  {author} {\bibinfo {author} {\bibfnamefont {O.}~\bibnamefont {{Di
  Stefano}}}, \bibinfo {author} {\bibfnamefont {R.}~\bibnamefont {Stassi}},
  \bibinfo {author} {\bibfnamefont {L.}~\bibnamefont {Garziano}}, \bibinfo
  {author} {\bibfnamefont {A.~F.}\ \bibnamefont {Kockum}}, \bibinfo {author}
  {\bibfnamefont {S.}~\bibnamefont {Savasta}},\ and\ \bibinfo {author}
  {\bibfnamefont {F.}~\bibnamefont {Nori}},\ }\href
  {https://doi.org/10.1088/1367-2630/aa6cd7} {\bibfield  {journal} {\bibinfo
  {journal} {New J. Phys.}\ }\textbf {\bibinfo {volume} {19}},\ \bibinfo
  {pages} {053010} (\bibinfo {year} {2017})}\BibitemShut {NoStop}%
\bibitem [{\citenamefont {Macr{\`\i}}\ \emph {et~al.}(2018)\citenamefont
  {Macr{\`\i}}, \citenamefont {Nori},\ and\ \citenamefont
  {Kockum}}]{Macri2018a}%
  \BibitemOpen
  \bibfield  {author} {\bibinfo {author} {\bibfnamefont {V.}~\bibnamefont
  {Macr{\`\i}}}, \bibinfo {author} {\bibfnamefont {F.}~\bibnamefont {Nori}},\
  and\ \bibinfo {author} {\bibfnamefont {A.}~\bibnamefont {Kockum}},\ }\href
  {https://doi.org/https://doi.org/10.1103/PhysRevA.98.062327} {\bibfield
  {journal} {\bibinfo  {journal} {Phys. Rev. A}\ }\textbf {\bibinfo {volume}
  {98}},\ \bibinfo {pages} {062327} (\bibinfo {year} {2018})}\BibitemShut
  {NoStop}%
\bibitem [{\citenamefont {Zheng}\ \emph {et~al.}(2018)\citenamefont {Zheng},
  \citenamefont {L\"u}, \citenamefont {Bin}, \citenamefont {Zhan},
  \citenamefont {Li},\ and\ \citenamefont {Wu}}]{Zheng2018}%
  \BibitemOpen
  \bibfield  {author} {\bibinfo {author} {\bibfnamefont {L.-L.}\ \bibnamefont
  {Zheng}}, \bibinfo {author} {\bibfnamefont {X.-Y.}\ \bibnamefont {L\"u}},
  \bibinfo {author} {\bibfnamefont {Q.}~\bibnamefont {Bin}}, \bibinfo {author}
  {\bibfnamefont {Z.-M.}\ \bibnamefont {Zhan}}, \bibinfo {author}
  {\bibfnamefont {S.}~\bibnamefont {Li}},\ and\ \bibinfo {author}
  {\bibfnamefont {Y.}~\bibnamefont {Wu}},\ }\href
  {https://doi.org/https://doi.org/10.1103/PhysRevA.98.023863} {\bibfield
  {journal} {\bibinfo  {journal} {Phys. Rev. A}\ }\textbf {\bibinfo {volume}
  {98}},\ \bibinfo {pages} {023863} (\bibinfo {year} {2018})}\BibitemShut
  {NoStop}%
\bibitem [{\citenamefont {Felicetti}\ \emph {et~al.}(2018)\citenamefont
  {Felicetti}, \citenamefont {Rossatto}, \citenamefont {Rico}, \citenamefont
  {Solano},\ and\ \citenamefont {Forn-D\'{\i}az}}]{Felicetti2018}%
  \BibitemOpen
  \bibfield  {author} {\bibinfo {author} {\bibfnamefont {S.}~\bibnamefont
  {Felicetti}}, \bibinfo {author} {\bibfnamefont {D.~Z.}\ \bibnamefont
  {Rossatto}}, \bibinfo {author} {\bibfnamefont {E.}~\bibnamefont {Rico}},
  \bibinfo {author} {\bibfnamefont {E.}~\bibnamefont {Solano}},\ and\ \bibinfo
  {author} {\bibfnamefont {P.}~\bibnamefont {Forn-D\'{\i}az}},\ }\href
  {https://doi.org/https://doi.org/10.1103/PhysRevA.97.013851} {\bibfield
  {journal} {\bibinfo  {journal} {Phys. Rev. A}\ }\textbf {\bibinfo {volume}
  {97}},\ \bibinfo {pages} {013851} (\bibinfo {year} {2018})}\BibitemShut
  {NoStop}%
\bibitem [{\citenamefont {Macr{\`{i}}}\ \emph {et~al.}(2018)\citenamefont
  {Macr{\`{i}}}, \citenamefont {Ridolfo}, \citenamefont {{Di Stefano}},
  \citenamefont {Kockum}, \citenamefont {Nori},\ and\ \citenamefont
  {Savasta}}]{Macri2018}%
  \BibitemOpen
  \bibfield  {author} {\bibinfo {author} {\bibfnamefont {V.}~\bibnamefont
  {Macr{\`{i}}}}, \bibinfo {author} {\bibfnamefont {A.}~\bibnamefont
  {Ridolfo}}, \bibinfo {author} {\bibfnamefont {O.}~\bibnamefont {{Di
  Stefano}}}, \bibinfo {author} {\bibfnamefont {A.~F.}\ \bibnamefont {Kockum}},
  \bibinfo {author} {\bibfnamefont {F.}~\bibnamefont {Nori}},\ and\ \bibinfo
  {author} {\bibfnamefont {S.}~\bibnamefont {Savasta}},\ }\href
  {https://doi.org/10.1103/PhysRevX.8.011031} {\bibfield  {journal} {\bibinfo
  {journal} {Phys. Rev. X}\ }\textbf {\bibinfo {volume} {8}},\ \bibinfo {pages}
  {011031} (\bibinfo {year} {2018})}\BibitemShut {NoStop}%
\bibitem [{\citenamefont {Flick}\ \emph {et~al.}(2018)\citenamefont {Flick},
  \citenamefont {Sch\"{a}fer}, \citenamefont {Ruggenthaler}, \citenamefont
  {Appel},\ and\ \citenamefont {Rubio}}]{Flick2018}%
  \BibitemOpen
  \bibfield  {author} {\bibinfo {author} {\bibfnamefont {J.}~\bibnamefont
  {Flick}}, \bibinfo {author} {\bibfnamefont {C.}~\bibnamefont {Sch\"{a}fer}},
  \bibinfo {author} {\bibfnamefont {M.}~\bibnamefont {Ruggenthaler}}, \bibinfo
  {author} {\bibfnamefont {H.}~\bibnamefont {Appel}},\ and\ \bibinfo {author}
  {\bibfnamefont {A.}~\bibnamefont {Rubio}},\ }\href
  {https://doi.org/10.1021/acsphotonics.7b01279} {\bibfield  {journal}
  {\bibinfo  {journal} {ACS Photonics}\ }\textbf {\bibinfo {volume} {5}},\
  \bibinfo {pages} {992} (\bibinfo {year} {2018})}\BibitemShut {NoStop}%
\bibitem [{\citenamefont {S\'anchez-Burillo}\ \emph {et~al.}(2019)\citenamefont
  {S\'anchez-Burillo}, \citenamefont {Mart\'{\i}n-Moreno}, \citenamefont
  {Garc\'{\i}a-Ripoll},\ and\ \citenamefont {Zueco}}]{SanchezBurillo2019}%
  \BibitemOpen
  \bibfield  {author} {\bibinfo {author} {\bibfnamefont {E.}~\bibnamefont
  {S\'anchez-Burillo}}, \bibinfo {author} {\bibfnamefont {L.}~\bibnamefont
  {Mart\'{\i}n-Moreno}}, \bibinfo {author} {\bibfnamefont {J.~J.}\ \bibnamefont
  {Garc\'{\i}a-Ripoll}},\ and\ \bibinfo {author} {\bibfnamefont
  {D.}~\bibnamefont {Zueco}},\ }\href
  {https://doi.org/10.1103/PhysRevLett.123.013601} {\bibfield  {journal}
  {\bibinfo  {journal} {Phys. Rev. Lett.}\ }\textbf {\bibinfo {volume} {123}},\
  \bibinfo {pages} {013601} (\bibinfo {year} {2019})}\BibitemShut {NoStop}%
\bibitem [{\citenamefont {Mordovina}\ \emph {et~al.}(2019)\citenamefont
  {Mordovina}, \citenamefont {Bungey}, \citenamefont {Appel}, \citenamefont
  {Knowles}, \citenamefont {Rubio},\ and\ \citenamefont
  {Manby}}]{Mordovina2019}%
  \BibitemOpen
  \bibfield  {author} {\bibinfo {author} {\bibfnamefont {U.}~\bibnamefont
  {Mordovina}}, \bibinfo {author} {\bibfnamefont {C.}~\bibnamefont {Bungey}},
  \bibinfo {author} {\bibfnamefont {H.}~\bibnamefont {Appel}}, \bibinfo
  {author} {\bibfnamefont {P.~J.}\ \bibnamefont {Knowles}}, \bibinfo {author}
  {\bibfnamefont {A.}~\bibnamefont {Rubio}},\ and\ \bibinfo {author}
  {\bibfnamefont {F.~R.}\ \bibnamefont {Manby}},\ }\href
  {https://arxiv.org/abs/1909.02401} {\bibfield  {journal} {\bibinfo  {journal}
  {arXiv:1909.02401}\ } (\bibinfo {year} {2019})}\BibitemShut {NoStop}%
\bibitem [{\citenamefont {{Di Stefano}}\ \emph {et~al.}(2018)\citenamefont {{Di
  Stefano}}, \citenamefont {Kockum}, \citenamefont {Ridolfo}, \citenamefont
  {Savasta},\ and\ \citenamefont {Nori}}]{DiStefano2017a}%
  \BibitemOpen
  \bibfield  {author} {\bibinfo {author} {\bibfnamefont {O.}~\bibnamefont {{Di
  Stefano}}}, \bibinfo {author} {\bibfnamefont {A.~F.}\ \bibnamefont {Kockum}},
  \bibinfo {author} {\bibfnamefont {A.}~\bibnamefont {Ridolfo}}, \bibinfo
  {author} {\bibfnamefont {S.}~\bibnamefont {Savasta}},\ and\ \bibinfo {author}
  {\bibfnamefont {F.}~\bibnamefont {Nori}},\ }\href
  {https://doi.org/10.1038/s41598-018-36056-1} {\bibfield  {journal} {\bibinfo
  {journal} {Sci. Rep.}\ }\textbf {\bibinfo {volume} {8}},\ \bibinfo {pages}
  {17825} (\bibinfo {year} {2018})}\BibitemShut {NoStop}%
\bibitem [{\citenamefont {Law}(1995)}]{Law1995}%
  \BibitemOpen
  \bibfield  {author} {\bibinfo {author} {\bibfnamefont {C.~K.}\ \bibnamefont
  {Law}},\ }\href {https://doi.org/10.1103/PhysRevA.51.2537} {\bibfield
  {journal} {\bibinfo  {journal} {Phys. Rev. A}\ }\textbf {\bibinfo {volume}
  {51}},\ \bibinfo {pages} {2537} (\bibinfo {year} {1995})}\BibitemShut
  {NoStop}%
\bibitem [{\citenamefont {Di~Stefano}\ \emph {et~al.}(2019)\citenamefont
  {Di~Stefano}, \citenamefont {Settineri}, \citenamefont {Macr{\`\i}},
  \citenamefont {Ridolfo}, \citenamefont {Stassi}, \citenamefont {Kockum},
  \citenamefont {Savasta},\ and\ \citenamefont {Nori}}]{DiStefano2019prl}%
  \BibitemOpen
  \bibfield  {author} {\bibinfo {author} {\bibfnamefont {O.}~\bibnamefont
  {Di~Stefano}}, \bibinfo {author} {\bibfnamefont {A.}~\bibnamefont
  {Settineri}}, \bibinfo {author} {\bibfnamefont {V.}~\bibnamefont
  {Macr{\`\i}}}, \bibinfo {author} {\bibfnamefont {A.}~\bibnamefont {Ridolfo}},
  \bibinfo {author} {\bibfnamefont {R.}~\bibnamefont {Stassi}}, \bibinfo
  {author} {\bibfnamefont {A.~F.}\ \bibnamefont {Kockum}}, \bibinfo {author}
  {\bibfnamefont {S.}~\bibnamefont {Savasta}},\ and\ \bibinfo {author}
  {\bibfnamefont {F.}~\bibnamefont {Nori}},\ }\href
  {https://doi.org/https://doi.org/10.1103/PhysRevLett.122.030402} {\bibfield
  {journal} {\bibinfo  {journal} {Phys. Rev. Lett.}\ }\textbf {\bibinfo
  {volume} {122}},\ \bibinfo {pages} {030402} (\bibinfo {year}
  {2019})}\BibitemShut {NoStop}%
\bibitem [{\citenamefont {Kirton}\ \emph {et~al.}(2019)\citenamefont {Kirton},
  \citenamefont {Roses}, \citenamefont {Keeling},\ and\ \citenamefont
  {Dalla~Torre}}]{Kirton2019}%
  \BibitemOpen
  \bibfield  {author} {\bibinfo {author} {\bibfnamefont {P.}~\bibnamefont
  {Kirton}}, \bibinfo {author} {\bibfnamefont {M.~M.}\ \bibnamefont {Roses}},
  \bibinfo {author} {\bibfnamefont {J.}~\bibnamefont {Keeling}},\ and\ \bibinfo
  {author} {\bibfnamefont {E.~G.}\ \bibnamefont {Dalla~Torre}},\ }\href
  {https://onlinelibrary.wiley.com/doi/full/10.1002/qute.201800043?casa_token=fHpH24Oc00UAAAAA%3A3t_HL4M6vIyRF80_jpU8NBe1bdF6UliVLL1wRBFSgv61IXZ2uNku52ujWqAUmvaKPPJSG4f2NG8ZAu7QLQ}
  {\bibfield  {journal} {\bibinfo  {journal} {Adv. Quantum Technol.}\ }\textbf
  {\bibinfo {volume} {2}},\ \bibinfo {pages} {1800043} (\bibinfo {year}
  {2019})}\BibitemShut {NoStop}%
\bibitem [{\citenamefont {Flick}\ \emph {et~al.}(2017)\citenamefont {Flick},
  \citenamefont {Ruggenthaler}, \citenamefont {Appel},\ and\ \citenamefont
  {Rubio}}]{Flick2017}%
  \BibitemOpen
  \bibfield  {author} {\bibinfo {author} {\bibfnamefont {J.}~\bibnamefont
  {Flick}}, \bibinfo {author} {\bibfnamefont {M.}~\bibnamefont {Ruggenthaler}},
  \bibinfo {author} {\bibfnamefont {H.}~\bibnamefont {Appel}},\ and\ \bibinfo
  {author} {\bibfnamefont {A.}~\bibnamefont {Rubio}},\ }\href
  {https://www.pnas.org/content/114/12/3026} {\bibfield  {journal} {\bibinfo
  {journal} {Proc. Natl. Acad. Sci. U.S.A.}\ }\textbf {\bibinfo {volume}
  {114}},\ \bibinfo {pages} {3026} (\bibinfo {year} {2017})}\BibitemShut
  {NoStop}%
\bibitem [{\citenamefont {Sakurai}(1994)}]{Sakurai1994}%
  \BibitemOpen
  \bibfield  {author} {\bibinfo {author} {\bibfnamefont {J.~J.}\ \bibnamefont
  {Sakurai}},\ }\href@noop {} {\emph {\bibinfo {title} {{Modern Quantum
  Mechanics}}}}\ (\bibinfo  {publisher} {Addison-Wesley Publishing Company,
  Inc.},\ \bibinfo {year} {1994})\BibitemShut {NoStop}%
\bibitem [{\citenamefont {{Di Stefano}}\ \emph {et~al.}(2019)\citenamefont {{Di
  Stefano}}, \citenamefont {Settineri}, \citenamefont {Macr{\`{i}}},
  \citenamefont {Garziano}, \citenamefont {Stassi}, \citenamefont {Savasta},\
  and\ \citenamefont {Nori}}]{DiStefano2019}%
  \BibitemOpen
  \bibfield  {author} {\bibinfo {author} {\bibfnamefont {O.}~\bibnamefont {{Di
  Stefano}}}, \bibinfo {author} {\bibfnamefont {A.}~\bibnamefont {Settineri}},
  \bibinfo {author} {\bibfnamefont {V.}~\bibnamefont {Macr{\`{i}}}}, \bibinfo
  {author} {\bibfnamefont {L.}~\bibnamefont {Garziano}}, \bibinfo {author}
  {\bibfnamefont {R.}~\bibnamefont {Stassi}}, \bibinfo {author} {\bibfnamefont
  {S.}~\bibnamefont {Savasta}},\ and\ \bibinfo {author} {\bibfnamefont
  {F.}~\bibnamefont {Nori}},\ }\href
  {https://doi.org/10.1038/s41567-019-0534-4} {\bibfield  {journal} {\bibinfo
  {journal} {Nat. Phys.}\ }\textbf {\bibinfo {volume} {15}},\ \bibinfo {pages}
  {803} (\bibinfo {year} {2019})}\BibitemShut {NoStop}%
\bibitem [{\citenamefont {{De Bernardis}}\ \emph
  {et~al.}(2018{\natexlab{a}})\citenamefont {{De Bernardis}}, \citenamefont
  {Pilar}, \citenamefont {Jaako}, \citenamefont {{De Liberato}},\ and\
  \citenamefont {Rabl}}]{DeBernardis2018}%
  \BibitemOpen
  \bibfield  {author} {\bibinfo {author} {\bibfnamefont {D.}~\bibnamefont {{De
  Bernardis}}}, \bibinfo {author} {\bibfnamefont {P.}~\bibnamefont {Pilar}},
  \bibinfo {author} {\bibfnamefont {T.}~\bibnamefont {Jaako}}, \bibinfo
  {author} {\bibfnamefont {S.}~\bibnamefont {{De Liberato}}},\ and\ \bibinfo
  {author} {\bibfnamefont {P.}~\bibnamefont {Rabl}},\ }\href
  {https://doi.org/10.1103/PhysRevA.98.053819} {\bibfield  {journal} {\bibinfo
  {journal} {Phys. Rev. A}\ }\textbf {\bibinfo {volume} {98}},\ \bibinfo
  {pages} {053819} (\bibinfo {year} {2018}{\natexlab{a}})}\BibitemShut
  {NoStop}%
\bibitem [{\citenamefont {Stokes}\ and\ \citenamefont
  {Nazir}(2019)}]{Stokes2018}%
  \BibitemOpen
  \bibfield  {author} {\bibinfo {author} {\bibfnamefont {A.}~\bibnamefont
  {Stokes}}\ and\ \bibinfo {author} {\bibfnamefont {A.}~\bibnamefont {Nazir}},\
  }\href {https://www.nature.com/articles/s41467-018-08101-0} {\bibfield
  {journal} {\bibinfo  {journal} {Nat. Commun.}\ }\textbf {\bibinfo {volume}
  {10}} (\bibinfo {year} {2019})}\BibitemShut {NoStop}%
\bibitem [{\citenamefont {Settineri}\ \emph {et~al.}(2019)\citenamefont
  {Settineri}, \citenamefont {Di~Stefano}, \citenamefont {Zueco}, \citenamefont
  {Hughes}, \citenamefont {Savasta},\ and\ \citenamefont {F.}}]{Settineri2019}%
  \BibitemOpen
  \bibfield  {author} {\bibinfo {author} {\bibfnamefont {A.}~\bibnamefont
  {Settineri}}, \bibinfo {author} {\bibfnamefont {O.}~\bibnamefont
  {Di~Stefano}}, \bibinfo {author} {\bibfnamefont {D.}~\bibnamefont {Zueco}},
  \bibinfo {author} {\bibfnamefont {S.}~\bibnamefont {Hughes}}, \bibinfo
  {author} {\bibfnamefont {S.}~\bibnamefont {Savasta}},\ and\ \bibinfo {author}
  {\bibfnamefont {N.}~\bibnamefont {F.}},\ }\href
  {https://arxiv.org/abs/1912.08548} {\bibfield  {journal} {\bibinfo  {journal}
  {arXiv:1912.08548}\ } (\bibinfo {year} {2019})}\BibitemShut {NoStop}%
\bibitem [{\citenamefont {Jacobs}\ and\ \citenamefont
  {Landahl}(2009)}]{Jacobs2009}%
  \BibitemOpen
  \bibfield  {author} {\bibinfo {author} {\bibfnamefont {K.}~\bibnamefont
  {Jacobs}}\ and\ \bibinfo {author} {\bibfnamefont {A.~J.}\ \bibnamefont
  {Landahl}},\ }\href {https://doi.org/10.1103/PhysRevLett.103.067201}
  {\bibfield  {journal} {\bibinfo  {journal} {Phys. Rev. Lett.}\ }\textbf
  {\bibinfo {volume} {103}},\ \bibinfo {pages} {067201} (\bibinfo {year}
  {2009})}\BibitemShut {NoStop}%
\bibitem [{\citenamefont {Ferretti}\ and\ \citenamefont
  {Gerace}(2012)}]{Ferretti2012}%
  \BibitemOpen
  \bibfield  {author} {\bibinfo {author} {\bibfnamefont {S.}~\bibnamefont
  {Ferretti}}\ and\ \bibinfo {author} {\bibfnamefont {D.}~\bibnamefont
  {Gerace}},\ }\href {https://doi.org/10.1103/PhysRevB.85.033303} {\bibfield
  {journal} {\bibinfo  {journal} {Phys. Rev. B}\ }\textbf {\bibinfo {volume}
  {85}},\ \bibinfo {pages} {033303} (\bibinfo {year} {2012})}\BibitemShut
  {NoStop}%
\bibitem [{\citenamefont {Malekakhlagh}\ and\ \citenamefont
  {T{\"{u}}reci}(2016)}]{Malekakhlagh2016}%
  \BibitemOpen
  \bibfield  {author} {\bibinfo {author} {\bibfnamefont {M.}~\bibnamefont
  {Malekakhlagh}}\ and\ \bibinfo {author} {\bibfnamefont {H.~E.}\ \bibnamefont
  {T{\"{u}}reci}},\ }\href {https://doi.org/10.1103/PhysRevA.93.012120}
  {\bibfield  {journal} {\bibinfo  {journal} {Phys. Rev. A}\ }\textbf {\bibinfo
  {volume} {93}},\ \bibinfo {pages} {012120} (\bibinfo {year}
  {2016})}\BibitemShut {NoStop}%
\bibitem [{\citenamefont {{S{\'{a}}nchez Mu{\~{n}}oz}}\ \emph
  {et~al.}(2018)\citenamefont {{S{\'{a}}nchez Mu{\~{n}}oz}}, \citenamefont
  {Nori},\ and\ \citenamefont {{De Liberato}}}]{SanchezMunoz2018}%
  \BibitemOpen
  \bibfield  {author} {\bibinfo {author} {\bibfnamefont {C.}~\bibnamefont
  {{S{\'{a}}nchez Mu{\~{n}}oz}}}, \bibinfo {author} {\bibfnamefont
  {F.}~\bibnamefont {Nori}},\ and\ \bibinfo {author} {\bibfnamefont
  {S.}~\bibnamefont {{De Liberato}}},\ }\href
  {https://doi.org/10.1038/s41467-018-04339-w} {\bibfield  {journal} {\bibinfo
  {journal} {Nat. Commun.}\ }\textbf {\bibinfo {volume} {9}},\ \bibinfo {pages}
  {1924} (\bibinfo {year} {2018})},\ \Eprint {https://arxiv.org/abs/1709.09872}
  {arXiv:1709.09872} \BibitemShut {NoStop}%
\bibitem [{\citenamefont {Kn{\"u}ppel}\ \emph {et~al.}(2019)\citenamefont
  {Kn{\"u}ppel}, \citenamefont {Ravets}, \citenamefont {Kroner}, \citenamefont
  {F{\"a}lt}, \citenamefont {Wegscheider},\ and\ \citenamefont
  {Imamoglu}}]{Knuppel2019}%
  \BibitemOpen
  \bibfield  {author} {\bibinfo {author} {\bibfnamefont {P.}~\bibnamefont
  {Kn{\"u}ppel}}, \bibinfo {author} {\bibfnamefont {S.}~\bibnamefont {Ravets}},
  \bibinfo {author} {\bibfnamefont {M.}~\bibnamefont {Kroner}}, \bibinfo
  {author} {\bibfnamefont {S.}~\bibnamefont {F{\"a}lt}}, \bibinfo {author}
  {\bibfnamefont {W.}~\bibnamefont {Wegscheider}},\ and\ \bibinfo {author}
  {\bibfnamefont {A.}~\bibnamefont {Imamoglu}},\ }\href@noop {} {\bibfield
  {journal} {\bibinfo  {journal} {Nature}\ }\textbf {\bibinfo {volume} {572}},\
  \bibinfo {pages} {91} (\bibinfo {year} {2019})}\BibitemShut {NoStop}%
\bibitem [{\citenamefont {{De Bernardis}}\ \emph
  {et~al.}(2018{\natexlab{b}})\citenamefont {{De Bernardis}}, \citenamefont
  {Jaako},\ and\ \citenamefont {Rabl}}]{DeBernardis2018a}%
  \BibitemOpen
  \bibfield  {author} {\bibinfo {author} {\bibfnamefont {D.}~\bibnamefont {{De
  Bernardis}}}, \bibinfo {author} {\bibfnamefont {T.}~\bibnamefont {Jaako}},\
  and\ \bibinfo {author} {\bibfnamefont {P.}~\bibnamefont {Rabl}},\ }\href
  {https://doi.org/10.1103/PhysRevA.97.043820} {\bibfield  {journal} {\bibinfo
  {journal} {Phys. Rev. A}\ }\textbf {\bibinfo {volume} {97}},\ \bibinfo
  {pages} {043820} (\bibinfo {year} {2018}{\natexlab{b}})}\BibitemShut
  {NoStop}%
\bibitem [{\citenamefont {Bin}\ \emph {et~al.}(2019)\citenamefont {Bin},
  \citenamefont {L\"u}, \citenamefont {Yin}, \citenamefont {Li},\ and\
  \citenamefont {Wu}}]{Bin2019}%
  \BibitemOpen
  \bibfield  {author} {\bibinfo {author} {\bibfnamefont {Q.}~\bibnamefont
  {Bin}}, \bibinfo {author} {\bibfnamefont {X.}~\bibnamefont {L\"u}}, \bibinfo
  {author} {\bibfnamefont {T.}~\bibnamefont {Yin}}, \bibinfo {author}
  {\bibfnamefont {Y.}~\bibnamefont {Li}},\ and\ \bibinfo {author}
  {\bibfnamefont {Y.}~\bibnamefont {Wu}},\ }\href
  {https://doi.org/https://doi.org/10.1103/PhysRevA.99.033809} {\bibfield
  {journal} {\bibinfo  {journal} {Phys. Rev. A}\ }\textbf {\bibinfo {volume}
  {99}},\ \bibinfo {pages} {033809} (\bibinfo {year} {2019})}\BibitemShut
  {NoStop}%
\bibitem [{\citenamefont {Aspelmeyer}\ \emph {et~al.}(2014)\citenamefont
  {Aspelmeyer}, \citenamefont {Kippenberg},\ and\ \citenamefont
  {Marquardt}}]{Aspelmeyer2014}%
  \BibitemOpen
  \bibfield  {author} {\bibinfo {author} {\bibfnamefont {M.}~\bibnamefont
  {Aspelmeyer}}, \bibinfo {author} {\bibfnamefont {T.~J.}\ \bibnamefont
  {Kippenberg}},\ and\ \bibinfo {author} {\bibfnamefont {F.}~\bibnamefont
  {Marquardt}},\ }\href {https://doi.org/10.1103/RevModPhys.86.1391} {\bibfield
   {journal} {\bibinfo  {journal} {Rev. Mod. Phys.}\ }\textbf {\bibinfo
  {volume} {86}},\ \bibinfo {pages} {1391} (\bibinfo {year}
  {2014})}\BibitemShut {NoStop}%
\bibitem [{\citenamefont {Beaudoin}\ \emph {et~al.}(2011)\citenamefont
  {Beaudoin}, \citenamefont {Gambetta},\ and\ \citenamefont
  {Blais}}]{Beaudoin2011}%
  \BibitemOpen
  \bibfield  {author} {\bibinfo {author} {\bibfnamefont {F.}~\bibnamefont
  {Beaudoin}}, \bibinfo {author} {\bibfnamefont {J.~M.}\ \bibnamefont
  {Gambetta}},\ and\ \bibinfo {author} {\bibfnamefont {A.}~\bibnamefont
  {Blais}},\ }\href {https://doi.org/10.1103/PhysRevA.84.043832} {\bibfield
  {journal} {\bibinfo  {journal} {Phys. Rev. A}\ }\textbf {\bibinfo {volume}
  {84}},\ \bibinfo {pages} {043832} (\bibinfo {year} {2011})},\ \Eprint
  {https://arxiv.org/abs/1107.3990} {arXiv:1107.3990} \BibitemShut {NoStop}%
\bibitem [{\citenamefont {Settineri}\ \emph {et~al.}(2018)\citenamefont
  {Settineri}, \citenamefont {Macr{\`{i}}}, \citenamefont {Ridolfo},
  \citenamefont {Di~Stefano}, \citenamefont {Kockum}, \citenamefont {Nori},\
  and\ \citenamefont {Savasta}}]{Settineri2018}%
  \BibitemOpen
  \bibfield  {author} {\bibinfo {author} {\bibfnamefont {A.}~\bibnamefont
  {Settineri}}, \bibinfo {author} {\bibfnamefont {V.}~\bibnamefont
  {Macr{\`{i}}}}, \bibinfo {author} {\bibfnamefont {A.}~\bibnamefont
  {Ridolfo}}, \bibinfo {author} {\bibfnamefont {O.}~\bibnamefont {Di~Stefano}},
  \bibinfo {author} {\bibfnamefont {A.}~\bibnamefont {Kockum}}, \bibinfo
  {author} {\bibfnamefont {F.}~\bibnamefont {Nori}},\ and\ \bibinfo {author}
  {\bibfnamefont {S.}~\bibnamefont {Savasta}},\ }\href
  {https://doi.org/https://doi.org/10.1103/PhysRevA.98.053834} {\bibfield
  {journal} {\bibinfo  {journal} {Phys. Rev. A}\ }\textbf {\bibinfo {volume}
  {98}},\ \bibinfo {pages} {053834} (\bibinfo {year} {2018})}\BibitemShut
  {NoStop}%
\end{thebibliography}%



\end{document}